\documentclass{aa}  
\usepackage{graphicx}
\usepackage{txfonts}
\usepackage{natbib}

\usepackage{color}
\begin{document}
   \title{High-resolution HST/ACS images of detached shells around carbon stars}


   \author{H. Olofsson \inst{1,2}
    	\and
          M. Maercker \inst{2}
          \and
          K. Eriksson \inst{3}
          \and
          B. Gustafsson \inst{3}
          \and
          F. Sch{\"o}ier \inst{1}
          }

   \offprints{H. Olofsson}

   \institute{Onsala Space Observatory, Dept. of Radio and Space Science, Chalmers University of Technology, SE-43992 Onsala, Sweden\\
              \email{hans.olofsson@chalmers.se}
         \and
             Department of Astronomy, AlbaNova University Center, Stockholm University, SE--10691 Stockholm, Sweden
         \and
             Department of Physics and Astronomy, Uppsala University, SE--75120 Uppsala, Sweden   
             }

   \date{Received 21 December 2009; accepted 28 January 2010}

 
  \abstract
   {Overall spherically symmetric, geometrically thin gas and dust shells have been found around a handful of asymptotic giant branch (AGB) carbon stars. Their dynamical ages lie in the range of 10$^3$ to 10$^4$ years. A tentative explanation for their existence is that they have formed as a consequence of mass-loss-rate modulations during a He-shell flash.}
   {The detached shells carry information on their formation process, as well as on the small-scale structure of the circumstellar medium around AGB stars due to the absence of significant line-of-sight confusion.}
   {The youngest detached shells, those around the carbon stars R~Scl and U~Cam, are studied here in great detail in scattered stellar light with the Advanced Survey Camera on the Hubble Space Telescope. Quantitative results are derived assuming optically thin dust scattering.}
   {The detached dust shells around R~Scl and U~Cam are found to be consistent with an overall spherical symmetry. They have radii of 19$\farcs$2 (corresponding to a linear size of 8$\times$10$^{16}$\,cm) and 7$\farcs$7 (5$\times$10$^{16}$\,cm), widths of 1$\farcs$2 (5$\times$10$^{15}$\,cm) and 0$\farcs$6 (4$\times$10$^{15}$\,cm), and dust masses of 3$\times$10$^{-6}$ and 3$\times$10$^{-7}$\,$M_\odot$, respectively. The dynamical ages of the R~Scl and U~Cam shells are estimated to be 1700 and 700\,yr, respectively, and the shell widths correspond to time scales of 100 and 50\,yr, respectively. Small-scale structure in the form of less than arcsec-sized clumps is clearly seen in the images of the R~Scl shell. Average clump dust masses are estimated to be about 2$\times$10$^{-9}$\,$M_\odot$. Comparisons with CO line interferometer data show that the dust and gas shells coincide spatially, within the errors ($\le$\,1$\arcsec$ for U~Cam and $\approx$\,2$\arcsec$ for R~Scl).}
   {The results are consistent with the interpretation of geometrically thin gas and dust shells formed by a mass-loss eruption during a He-shell flash, and where interaction with a previous wind plays a role as well. The mass loss responsible for the shells must have been remarkably isotropic, and, if wind interaction plays a role, this also applies to the mass loss prior to the eruption. Clumpy structure is present in the R~Scl shell, possibly as a consequence of the mass loss itself, but more likely as a consequence of instabilities in the expanding shell.}


   \maketitle
%

\section{Introduction}

Mass loss from the surface is an important characteristic of stellar evolution
on the asymptotic giant branch (AGB). It is a common property of most M-type,
S-type, and all C-type AGB stars that has been
well established for decades  \citep{olofetal93a, ramsetal09}. Yet, many of its details, such as the mechanism behind it,
and its evolution with time and e.g. dependence on stellar mass, are essentially unknown
and remain the major obstacle for understanding stellar evolution on (and beyond)
the AGB in detail as well as the contribution that AGB stars make to the galactic chemical
evolution \citep{habi96, will00, schrsedl01}.

It is particularly important to understand the temporal evolution and the dependence
on direction of the stellar mass loss. The former determines to a large extent how the
star evolves, while the latter has a profound effect on the circumstellar evolution
beyond the AGB, e.g., the formation of planetary nebulae. Geometrically thin detached shells \citep{schoetal05b} are a phenomenon that bears on both issues. 
Some carbon stars, of which less than ten are known, show this phenomenon.
It has been suggested that these shells are the result of strong
mass-loss modulations during a thermal pulse \citep{olofetal90,schretal98,wachetal02} 
and that they are additionally affected by an interaction with the surrounding (relic) circumstellar envelope (CSE)
\citep{stefscho00, schoetal05b}. Recently, \citet{mattetal07} presented models where the response to structure variations during a thermal pulse of the dynamical atmosphere and the expanding gas and dust are studied in some considerable detail. Geometrically thin shells appear under certain circumstances as an effect of a mass-loss eruption and a subsequent interaction with a previous slower wind. Remarkably, these shells imply that at least during this
phase, the mass loss is very close to isotropic \citep{olofetal96, 
olofetal00, gonzetal03a, maeretal10}. The study of thin, detached shells also has a bearing on the small-scale structure
of the circumstellar medium, since the line-of-sight confusion is limited by the thinness
of the shells \citep{olofetal00}.

Most of the information on these detached shells stems from CO radio line observations. 
The first detections were made by \citet{olofetal88},  and to this date there are
seven known carbon stars with this type of shell \citep{schoetal05b}. \citet{lindetal99} and \citet{olofetal00} used the IRAM PdB mm-wave interferometer to show that the shells
are geometrically thin (width/radius $<$\,0.1) and remarkably spherically symmetric. \citet{schoetal05b} found 
evidence of effects of interacting winds, i.e., the shells are affected by their
progress in a previous slower stellar wind. \citet{gonzetal01} showed for the first
time that these shells could be imaged in stellar light scattered in the
circumstellar medium, and \citet{gonzetal03a} and \citet{maeretal10}, following up on
this using polarimetric imaging, showed that both dust and atoms act as
scattering agents. No M-type or S-type AGB star was found with similar
geomtrically thin shells, despite extensive searches \citep[see e.g.,][]{kersolof99, ramsetal09}.
Detached dust shells around AGB and post-AGB objects have been seen as well, but with much coarser resolution observations \citep{wateetal94, izumetal96, izumetal97, hashetal98, specetal00}.

We note here that some detached shells around AGB stars may have a
different origin. Neutral hydrogen 21\,cm observations of extended CSEs show that some AGB stars
are surrounded by large detached shells, whose presence is most likely due to
interactions between circumstellar winds of different epochs or an interaction
with the surrounding interstellar medium \citep{libeetal07}. Similar conclusions are
drawn based on Spitzer observations \citep{wareetal06}.

In the CO mm observations by \citet{olofetal96, olofetal00} it was found that the detached shells showed a clumpy structure. The optical observations by \citet{gonzetal01} verified this finding optically to some extent, but the angular resolution was severely limited by seeing. Moreover, scattering in the terrestrial atmosphere limited the studies of the shells close to the star. In order to increase the resolution and study the circumstellar envelope closer to the star, we decided to use the Hubble Space Telescope (HST). We present broadband filter images of the circumstellar environments
of the carbon stars R~Sculptoris (R~Scl) and U~Camelopardalis (U~Cam) obtained with
the Advanced Camera for Surveys (ACS) on the Hubble Space Telescope (HST).


\section{Observations}

The carbon stars R~Scl and U~Cam (their stellar properties are summarized in Table~\ref{t:stellar}) were observed with the ACS on the HST in November 2004 and
January 2005, respectively. The images were obtained in high-resolution mode through the
broadband filters f475 (not U~Cam), f606, and f814 centred at 4760\,{\AA} (width 1458\,{\AA}),
5907\,{\AA} (width 2342\,{\AA}), and 8333\,{\AA} (width 2511\,{\AA}), respectively,
using the coronographic option.
R~Scl, where the detached shell is extended (diameter $\approx$\,40$\arcsec$),
the large coronographic spot (3$\arcsec$ in diameter) was used to effectively block the light
of the bright star. For U~Cam the angular size of the shell is smaller (diameter $\approx$\,15$\arcsec$)
and the smaller coronographic spot (1$\farcs$8 in diameter) was used. R Scl was observed in two
rotation angles of the satellite (separated by 30$\degr$) in the f606 and f814 filters
to cover more of the shell. Finally, for each target star a nearby
reference star with similar magnitude and colour was observed in the same mode (HD1760 for R~Scl and HD20797 for U~Cam). 
These images were used to subtract the point spread function (psf) from the images
of the target sources. The exposure times are given in Table~\ref{t:sources}. In addition,
short exposures of 0.1\,s were obtained for the target stars without coronograph in all filters to provide
estimates of the stellar fluxes. 

\begin{table}
\caption[]{Stellar properties [data from \citet{schoetal05b}].}
\label{t:stellar}
$$ 
\begin{array}{p{0.25\linewidth}ccccccccc}
\hline
\hline
\noalign{\smallskip}
\multicolumn{1}{l}{\mathrm{Source}} &
\multicolumn{1}{c}{\mathrm{Var.\ type}} &&
\multicolumn{1}{c}{P} &&
\multicolumn{1}{c}{D} &&
\multicolumn{1}{c}{L_{\star}} &&
\multicolumn{1}{c}{T_{\star}}\\
&
&&
\multicolumn{1}{c}{\mathrm{[days]}} &&
\multicolumn{1}{c}{[\mathrm{pc}]} &&
\multicolumn{1}{c}{[\mathrm{L}_{\sun}]} &&
\multicolumn{1}{c}{[\mathrm{K}]} 
\\
\noalign{\smallskip}
\hline
\noalign{\smallskip}
R Scl        & \mathrm{SRb}&& 370 && 290 && 4300 && 2630 \\
U Cam      &  \mathrm{SRb} &&  400 && 430 && 7000 && 2700 \\
\noalign{\smallskip}
\hline
\end{array}
$$ 
\end{table}

\begin{table}
\begin{minipage}[t]{\columnwidth}
\caption{Observational parameters.}
\label{t:sources}
\centering
\begin{tabular}{lcccc}
\hline \hline
Source   & Filter  & $\lambda_{\rm c}$ & $\Delta \lambda$ & Exp.   \\
         &         & [{\AA}]          & [{\AA}]          & [$s$]\\
\hline
R Scl       & f475  &  4760 & 1458 & 1734\phantom{a}  \\
            & f606  &  5907 & 2342 & 1206$^{\mathrm{a}}$  \\
            & f814  &  8333 & 2511 & 1206$^{\mathrm{a}}$  \\
HD1760   & f475  &       &      & \phantom{1}820\phantom{a}  \\
            & f606  &       &      & \phantom{1}330\phantom{a} \\
            & f814  &       &      & \phantom{1}330\phantom{a}  \\
U Cam       & f606  &       &      & 1060\phantom{a}  \\
            & f814  &       &      & 1060\phantom{a}  \\
HD20797   & f606  &       &      & 1050\phantom{a}  \\
            & f814  &       &      & 1050\phantom{a}  \\
\hline
\end{tabular}
\begin{list}{}{}
\item[$^{\mathrm{a}}$] Observed with this exposure time in two rotation angles of the satellite
\end{list}
\end{minipage}
\end{table}


\section{Data reduction}

We used the bias-corrected, flat-fielded, and cosmic-ray-corrected images supplied by the
HST pipeline. For each image the background level was estimated and subtracted.
The images were rebinned to a pixel size of 0$\farcs$026.
At this point the relevant reference star images were subtracted from the target
star images. This was achieved by shifting in position and multiplying the flux scale
of the reference star image until the best result was obtained in a box (with a side
of 13$\arcsec$ and 6$\arcsec$ for R~Scl and U~Cam, respectively) centred on
the target star. Note that this subtraction procedure introduces some uncertainty in
the radial behaviour of the scattered light distribution.

\begin{figure*}[t]
   \centering
   \includegraphics[width=9cm]{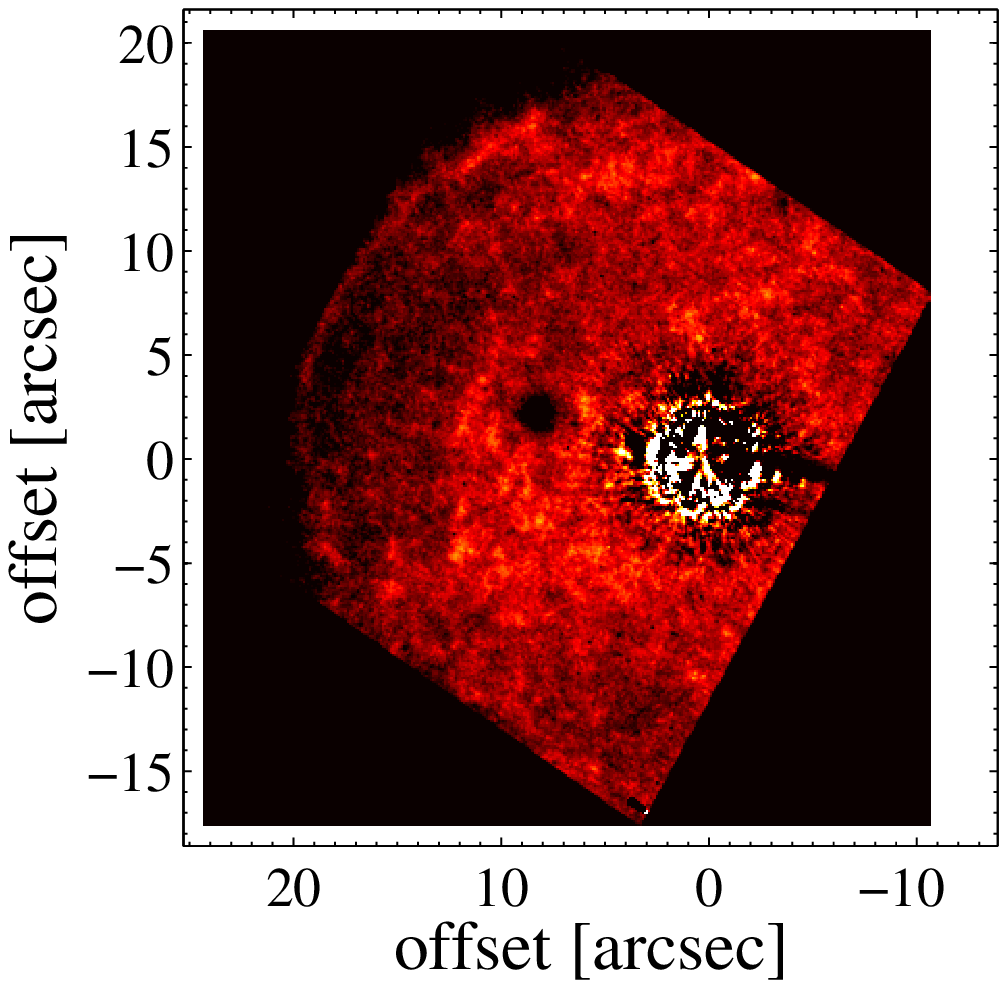}
   \includegraphics[width=9cm]{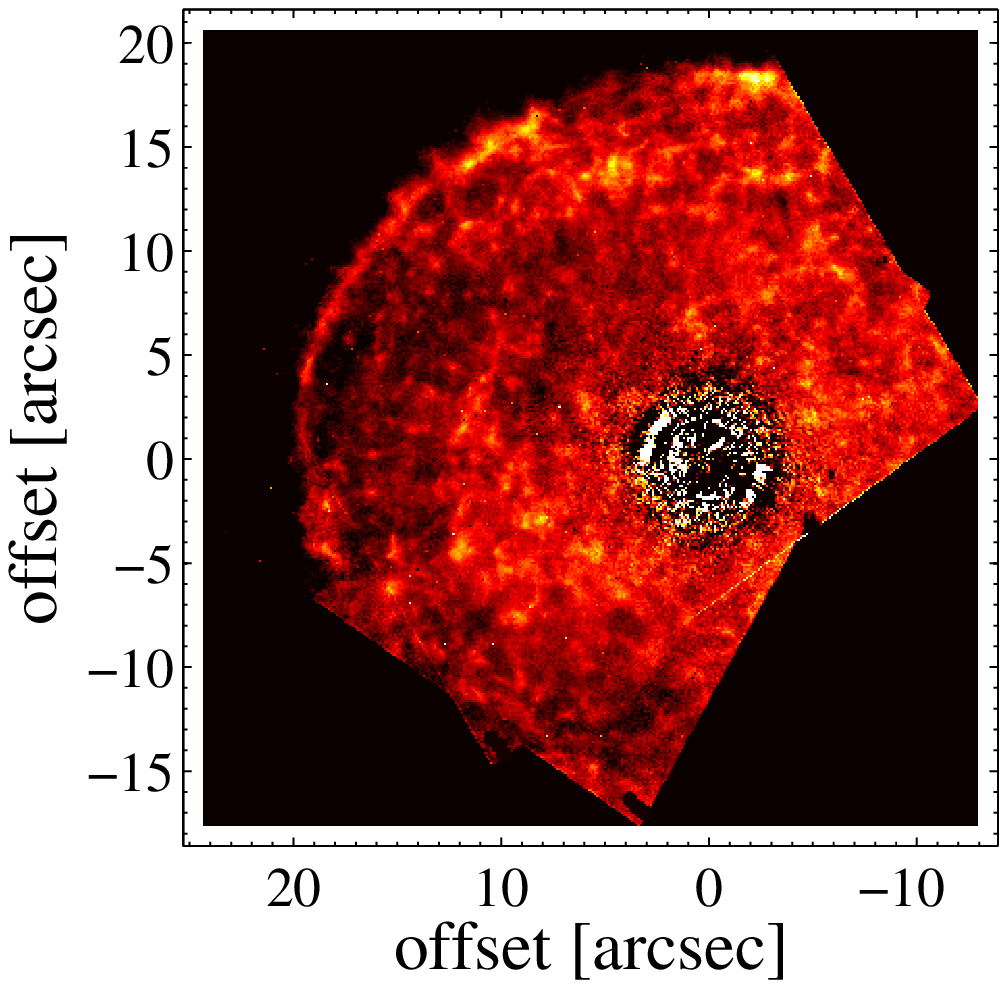}
   \includegraphics[width=9cm]{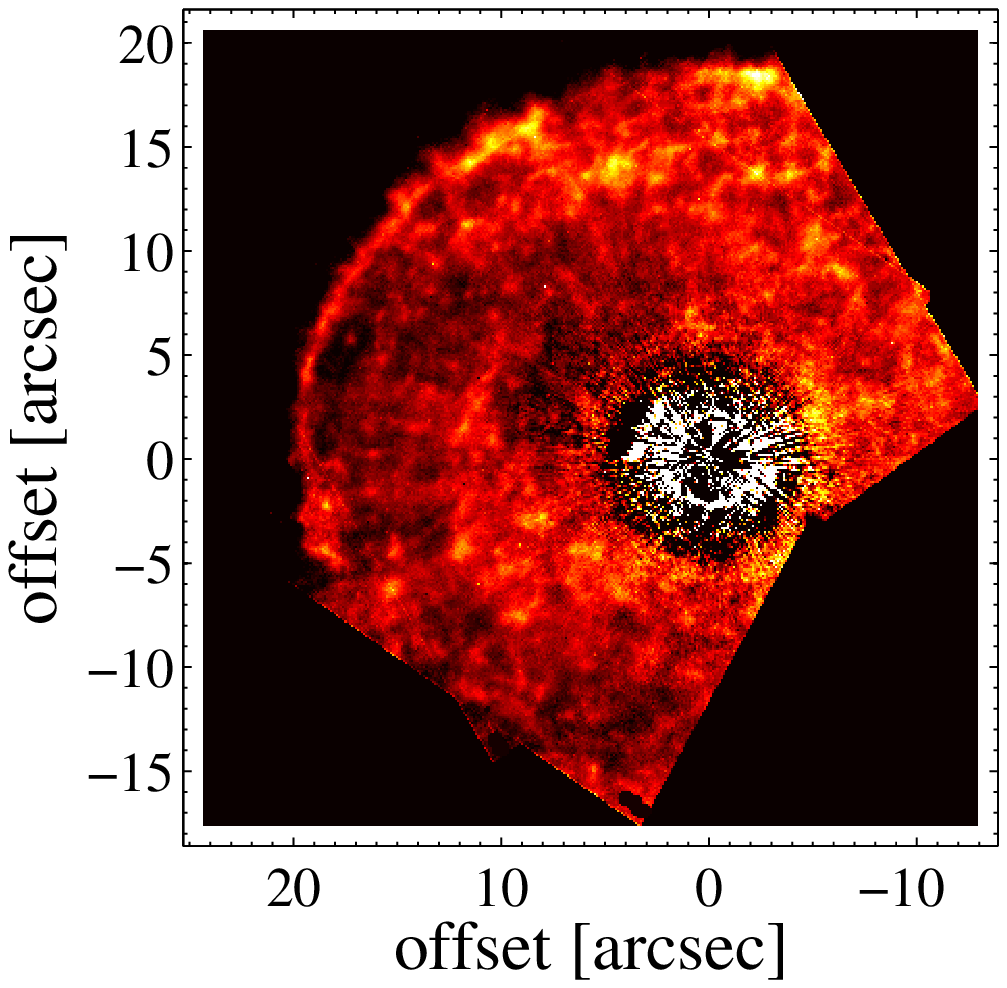}
   \includegraphics[width=9cm]{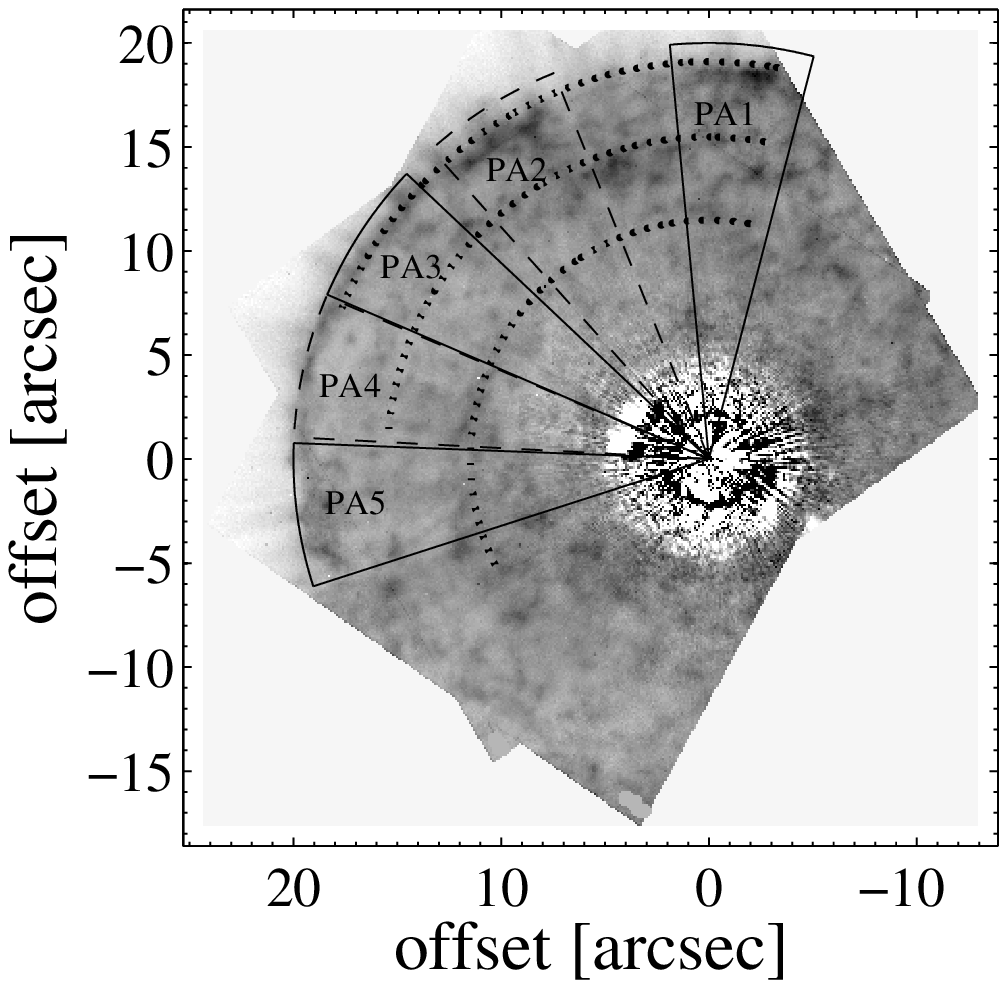}
   \caption{HST/ACS images of R~Scl in the f475 (top left), f606 (top right), and f814 (bottom left) filters. The size of the coronographic spot used is 3$\arcsec$. The 1$\farcs$8 coronographic spot and the coronographic `finger' is seen in the f475 image. About half of the shell is covered in these images. In the bottom right panel the sectors indicate the different PA intervals over which the AARPSs have been averaged. The centres of the sectors are all at the stellar position.}
   \label{f:rsclimages}
\end{figure*}

\begin{figure*}[t]
   \centering
   \includegraphics[width=9cm]{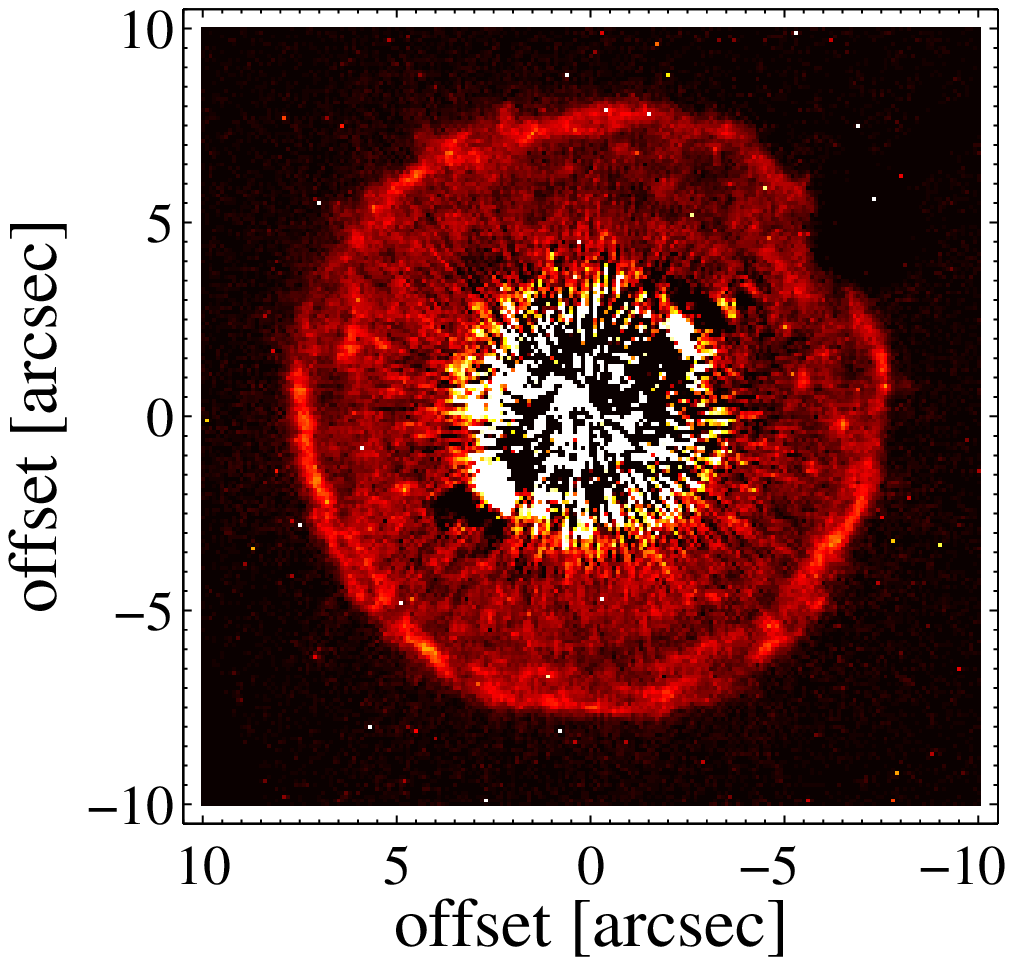}
   \includegraphics[width=9cm]{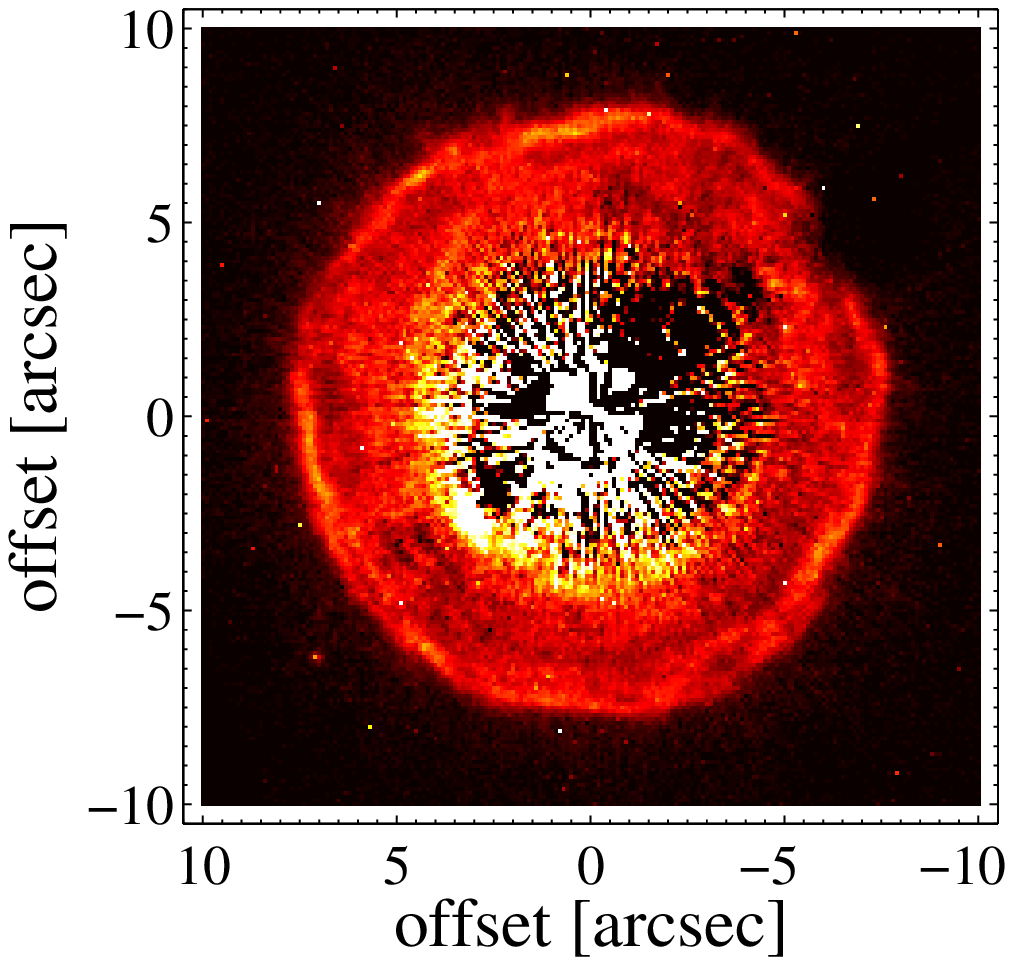}
   \includegraphics[width=9cm]{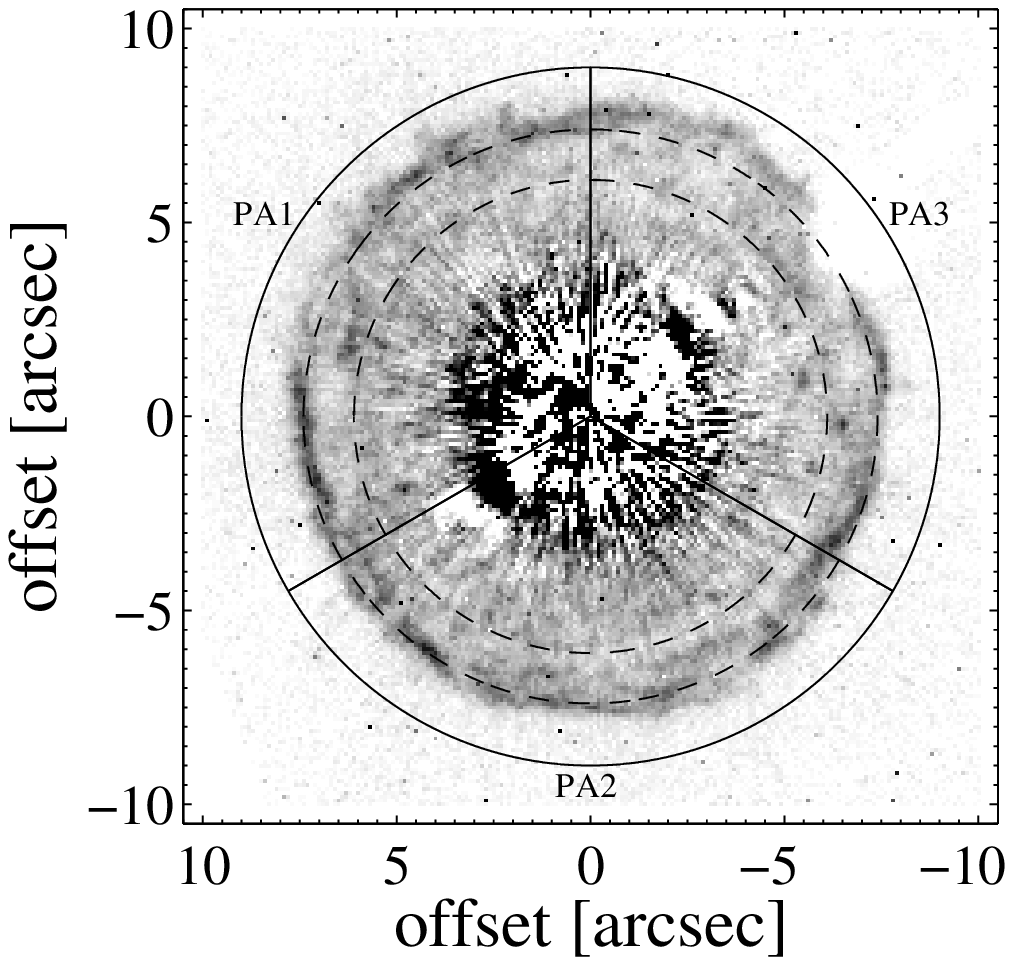}\hspace{90mm}
   \caption{HST/ACS images of U~Cam in the f606 (top left), and f814 (top right) filters, and the different PA intervals over which the brightness distributions were averaged (bottom left). The size of the coronographic spot used is 1$\farcs$8. The 3$\arcsec$ coronographic spot is seen at PA\,$\approx$\,$-45^\circ$ at the edge of the shell.}
   \label{f:ucamimages}
\end{figure*}

The resulting (psf-subtracted) images were geometrically corrected
and rotated to the ($\alpha,\delta$)-system using the correction files on the ACS homepage
and the relevant procedure in {\sc PYRAF}\footnote{http://www.stsci.edu/resources/software$\_$hardware/pyraf} (notably `pydrizzle'). The images of R~Scl obtained
at the two rotation angles of the satellite (only for
the f606 and f814 filters) were finally combined (the position of the
target star was estimated in both images and they were appropriately shifted in position
before combining). In this process the artifacts produced by the 1$\farcs$8 coronographic
spot and the coronographic `finger' were eliminated since these features appear
in different regions of the rotated images. Finally, all images were median filtered. 
The resulting images are shown in
Figs~\ref{f:rsclimages} and \ref{f:ucamimages}.

The stellar fluxes were obtained from 0.1\,s exposure
images. In the f475 filter the stellar flux density was obtained directly
from the image. For the f606 and f814 filters this procedure was not possible, 
since the area on the CCD which contains most of the stellar flux was saturated
even with 0.1\,s of exposure time. The stellar fluxes therefore were estimated 
through psf fitting. The publicly available program `Tiny Tim' was used to produce model HST psfs,
which were rotated and resampled to have the same rotation and pixel scale
as the observed images. This model psf image was fitted to the target image,
excluding the central saturated region, which made it possible to estimate the stellar flux densities in
both filters.

The presence of a circumstellar medium is apparent in all images. A comparison with
the data obtained for R~Scl by \citet{gonzetal01} using
EFOSC1 on the ESO 3.6\,m telescope is instructive. Clearly, the superior quality 
of the HST psf is crucial for this type of observations of extended low-surface-brightness
emission in the vicinity of a bright object. Only an area with a diameter $\approx$\,7$\arcsec$
around the central star is seriously contaminated in the HST images.

\section{Results}
\label{s:results}

\subsection{The nature of the circumstellar emission}

Circumstellar envelopes around AGB stars are cool objects (the kinetic and dust temperatures are $<$100\,K 
if we exclude the inner regions) and their emission, whether in lines or continuum,
is very weak at optical wavelengths. Therefore, the circumstellar emission seen in
our images must be due to scattered light. Further, it is scattered stellar light,
as opposed to e.g. the case of the extreme carbon star IRC+10216, where the external parts
of the CSE is seen through scattering of the interstellar radiation field
\citep[see e.g.,][]{maurhugg99}. These conclusions were firmly drawn
already by \citet{gonzetal01} in their analysis of the circumstellar
optical emission detected by them towards the carbon stars R~Scl and U~Ant.

The dominant scattering agent responsible for the emission in the images presented in this paper must be dust
particles for the following reason. The width of a circumstellar line from a 
carbon star is about 30\,km\,s$^{-1}$ (a well
established result from radio line observations; e.g., \citet{schoolof01}),
and this corresponds to $\approx$\,0.6\,{\AA} at 6000\,{\AA}. Consequently, a 
line is effectively diluted by a factor of about 4000 in the broadband filters 
used here. This does not absolutely exclude line scattering. However, as will be shown below,
for RÐScl the surface flux densities (i.e., flux per unit wavelength) in our 
images are consistent with those obtained by \citet{gonzetal01} in 50\,{\AA}
wide filters centred on the resonance lines of Na and K. Specifically, the flux densities
are the same in the f606 filter used by us and the 50\,{\AA} wide f59 filter centred on the NaD lines used by \citet{gonzetal01}, and therefore any contribution from the NaD line to the f606 image must be negligible. 


Nevertheless, scattering in the resonance lines of NaI and KI was observed in detached shells by \citet{maeretal10} and in ``normal'' circumstellar envelopes of carbon stars by e.g. \citet{gustetal97}, using spectroscopic techniques.

\begin{figure*}[t]
   \centering
      \includegraphics[width=6.3cm]{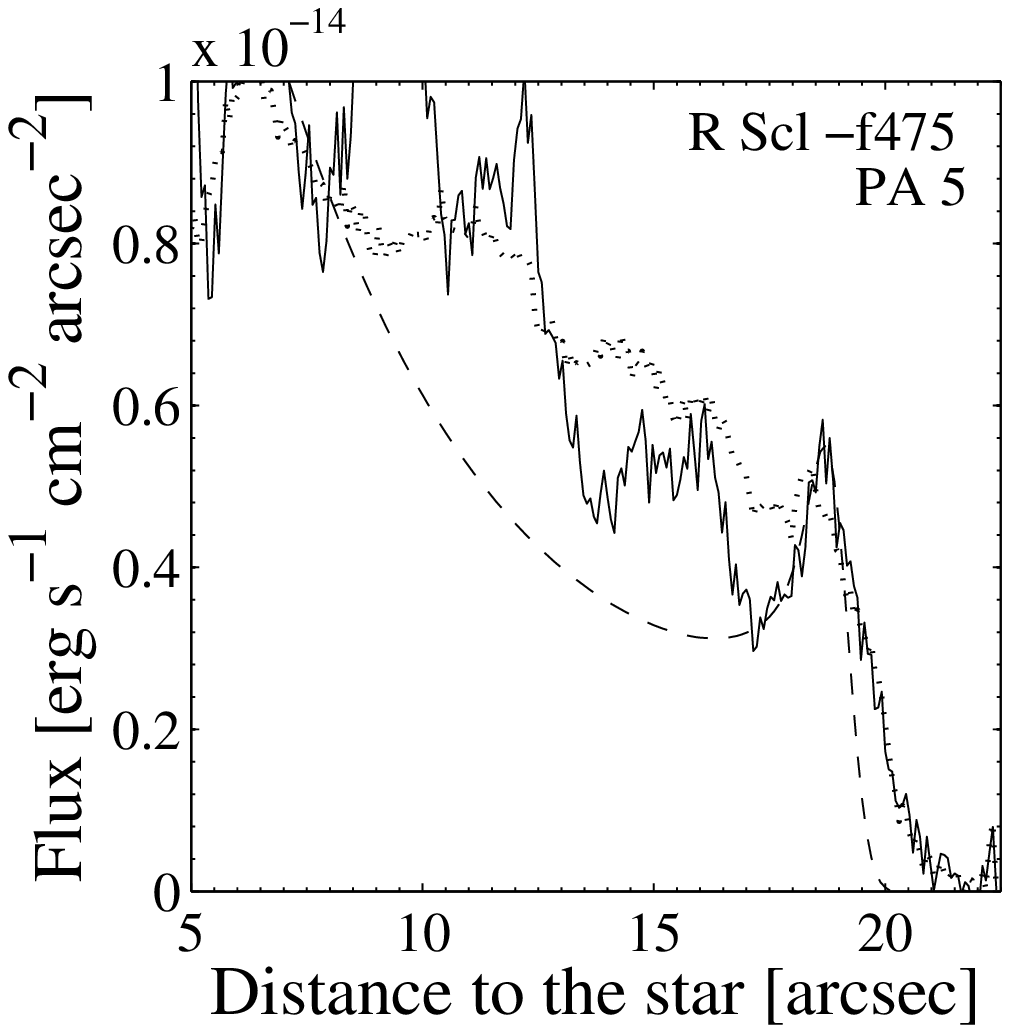}
      \includegraphics[width=6.3cm]{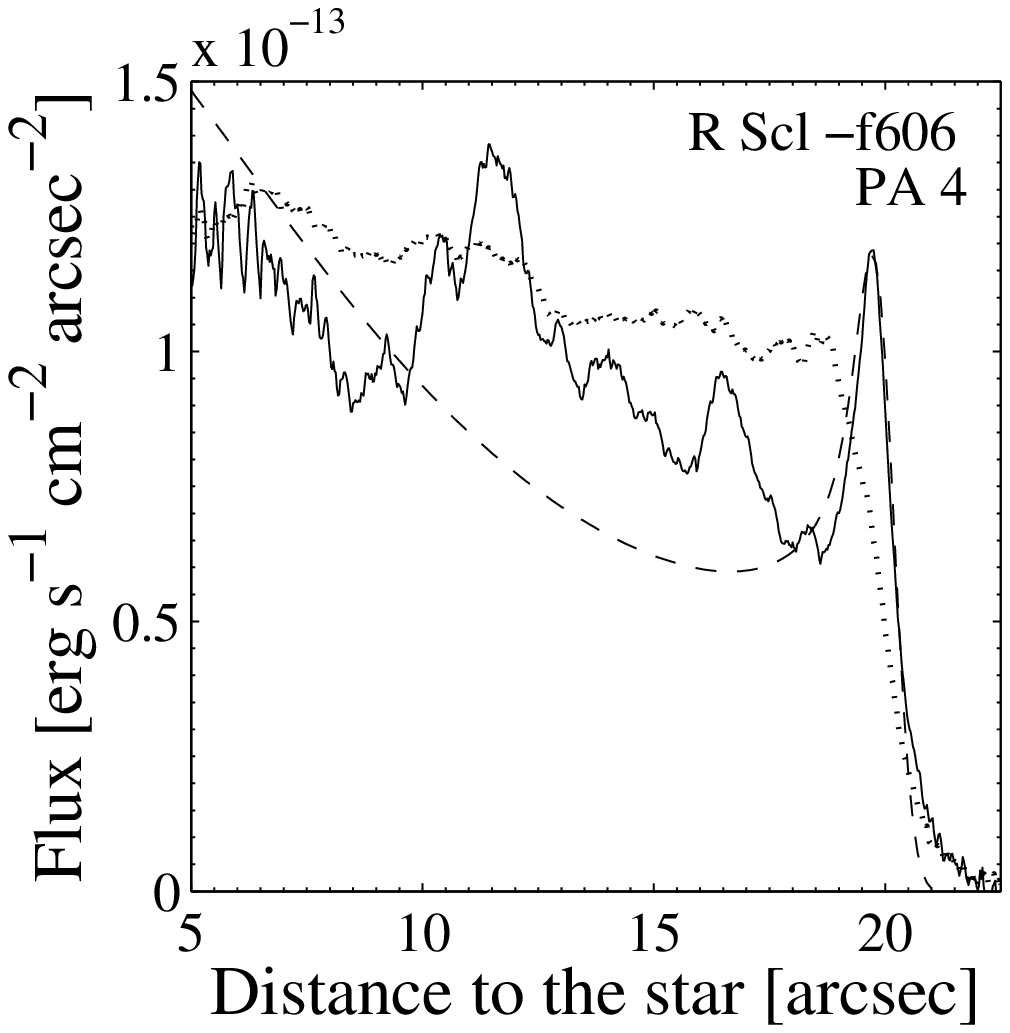}
      \includegraphics[width=6.3cm]{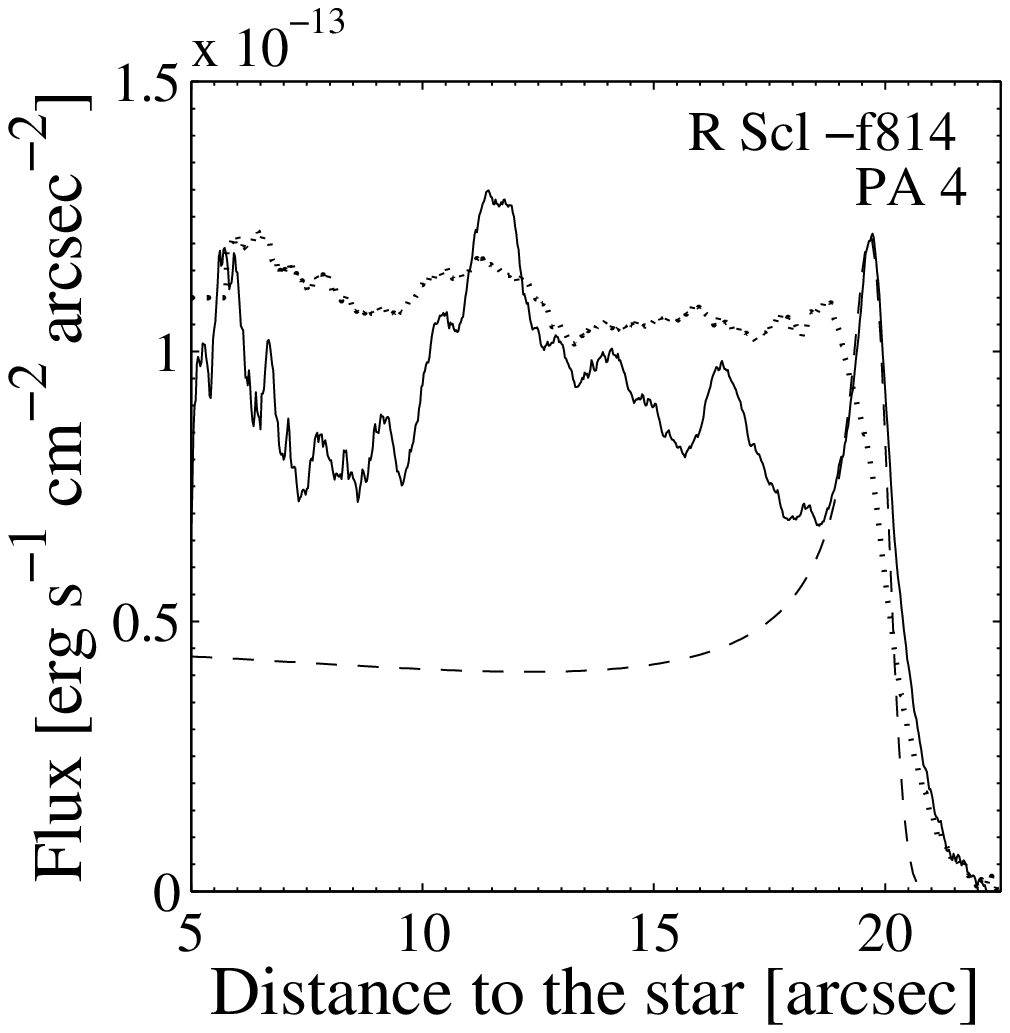}
   \caption{HST/ACS AARPs of R~Scl in the f475 (left), f606 (middle), and f814 (right) filters. The panels show the AARP in PA4 (solid line) (for the f475 filter PA5 is shown), the AARPs averaged over all PAs (dotted line), and fits assuming scattering of stellar light in a detached shell of dust (dashed line). See text for details.}
   \label{f:rsclaarps}
\end{figure*}

\subsection{The appearance of the circumstellar emission}
\label{rsclappear}

\subsubsection{R Scl}
\label{s:rsclappear}

Due to the large extent of the circumstellar emitting region around R~Scl compared to the field of view of
the ACS in high-resolution mode only a part of it could be covered in the observations, about 50\% of the area and about 30\% of the circumference. There is a marked limb-brightening, which follows an essentially circular arc, suggesting an overall spherically symmetric geometry. This is consistent with previous observations of R~Scl \citep{gonzetal01, gonzetal03a} and other detached shells. However, a circular arc fitted to the limb brightening does not have a centre at the stellar position (as is evident in Fig.~\ref{f:rsclimages}), but it is offset by $\approx$\,0$\farcs$7 at a position angle (PA) of about 160$\degr$. The same situation was found for the detached shell around TT~Cyg, where the shell centre is offset 1$\farcs$7 from the stellar position \citep{olofetal00}. This could be due to binary companions. We caution that for R~Scl we cover only about 30\% of the shell limb, and for this reason we use the stellar position as the centre of the shell in the following analysis. 

In order to analyse the circumstellar scattered light, azimuthally averaged radial profiles (AARPs) were made in five PA intervals (see Fig.~\ref{f:rsclimages}). The PA intervals are centred on 355$^{\circ}$~(PA1), 32$^{\circ}$ (PA2), 57$^{\circ}$ (PA3), 77$^{\circ}$ (PA4), and 98$^{\circ}$ (PA5), and cover a range of $\pm10^{\circ}$. Figure~\ref{f:rsclaarps} shows the R~Scl AARPs of the brightness distribution in PA4 in the f606 and f814 filters, and PA5 in the f475 filter (PA4 is not entirely covered in the f475 filter). Also shown is the average AARP over the entire image for each filter. Overall, the surface brightness in the AARP covering the entire image is uniform out to a rather sharp outer edge.
A low scattering optical depth means that the front and the rear parts of the CSE contribute
equally to the resulting brightness distribution (assuming isotropic scattering). The amount of scattered light is much lower in the f474 filter due to
the marked drop in the light of carbon stars at shorter wavelengths. The
discussion here is therefore centred around the f606 and f814 images.

The limb brightening is apparent in the AARPs, and it is consistent with a CSE that has the form of a geometrically thin shell, and is in accordance with the polarimetric imaging of \citet{gonzetal03a}, which clearly shows the presence of a detached shell. The limb brightening is much more prominent in the AARP of PA4 compared to the total AARP, i.e., it varies in strength with the PA, partly due to an actual strength variation but also because the size of the shell varies somewhat with the PA (mainly an effect of the shell not being centred on the star). There is weak emission extending at least 3$\arcsec$ beyond the peak of the shell emission. 
The impression is that this is scattered light in material that protrudes in the radial
direction from the shell, but the limited field of view of the ACS in high resolution
mode combined with the large extent of the shell means that this area is poorly
covered in the observations. There are in addition artifical radial structures, `striations' due to the coronographic mask, at low surface brightness.

Looking in more detail at the brightness distributions reveals considerable patchiness, with a structure that is very similar in the f606 and f814 images. Indeed, in an image formed by the ratio of these two images there is no discernible patchiness, only noise around the average ratio. The same result is obtained when ratios involving the f475 image are produced. This also shows that there are very little colour variations over the shell in the scattered light, except that the f475/f606 and f606/f814 ratio images show a systematic trend of the emission becoming redder with increasing distance from the star, most likely an effect of the wavelength dependence of the scattering asymmetry (see below). However, we caution that the reference star subtraction may affect any radial distribution. There appears to be no large-scale dominant pattern, but both the front and the rear part of the shell contribute to the
scattered light, which complicates matter. The patchiness is only a weak modulation in the AARP averaged over all PAs, the peak-to-peak value of these variations being less than $\pm$5\%, while the patches affect the AARPs more at the separate PA intervals. The relevance of this for the detailed density distribution in the shell is discussed further in Sect.~\ref{patches}.

\begin{figure}
   \centering
   \includegraphics[width=6.5cm]{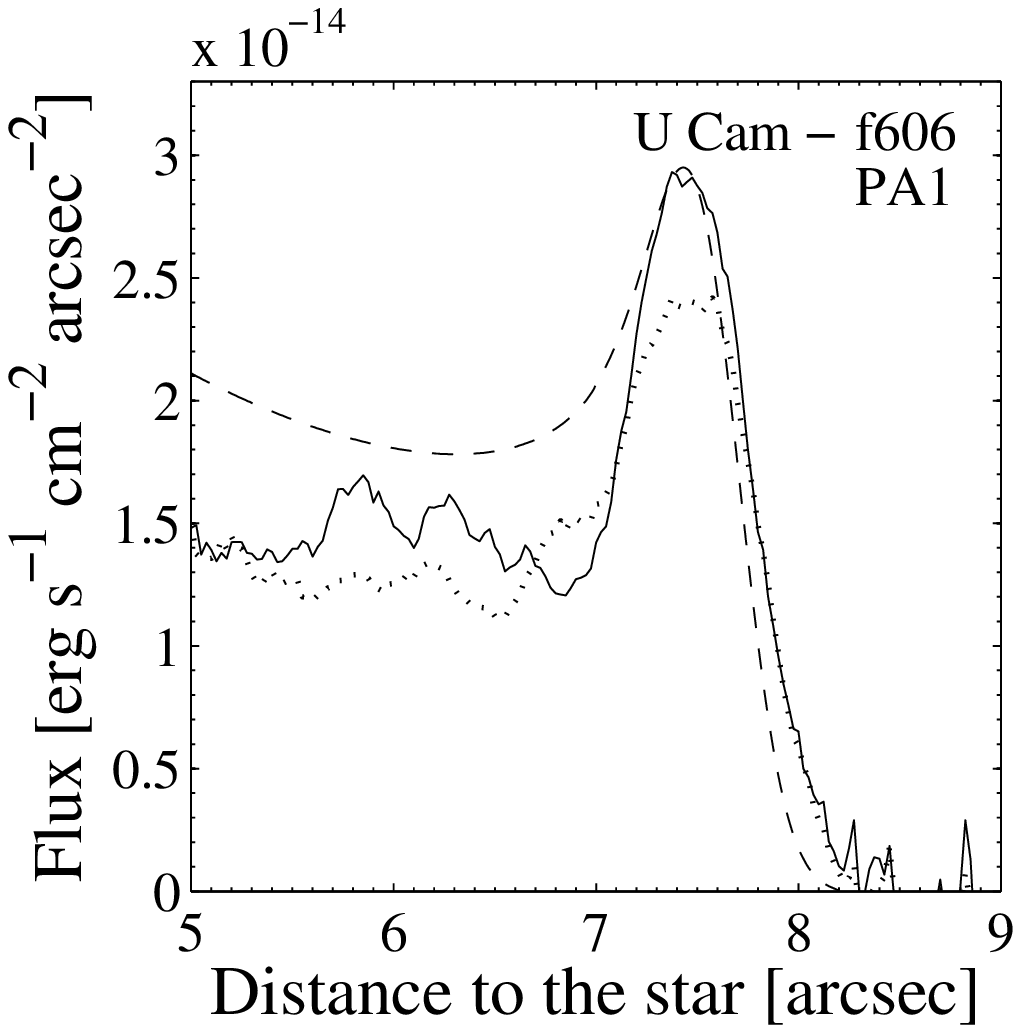}\\
   \includegraphics[width=6.5cm]{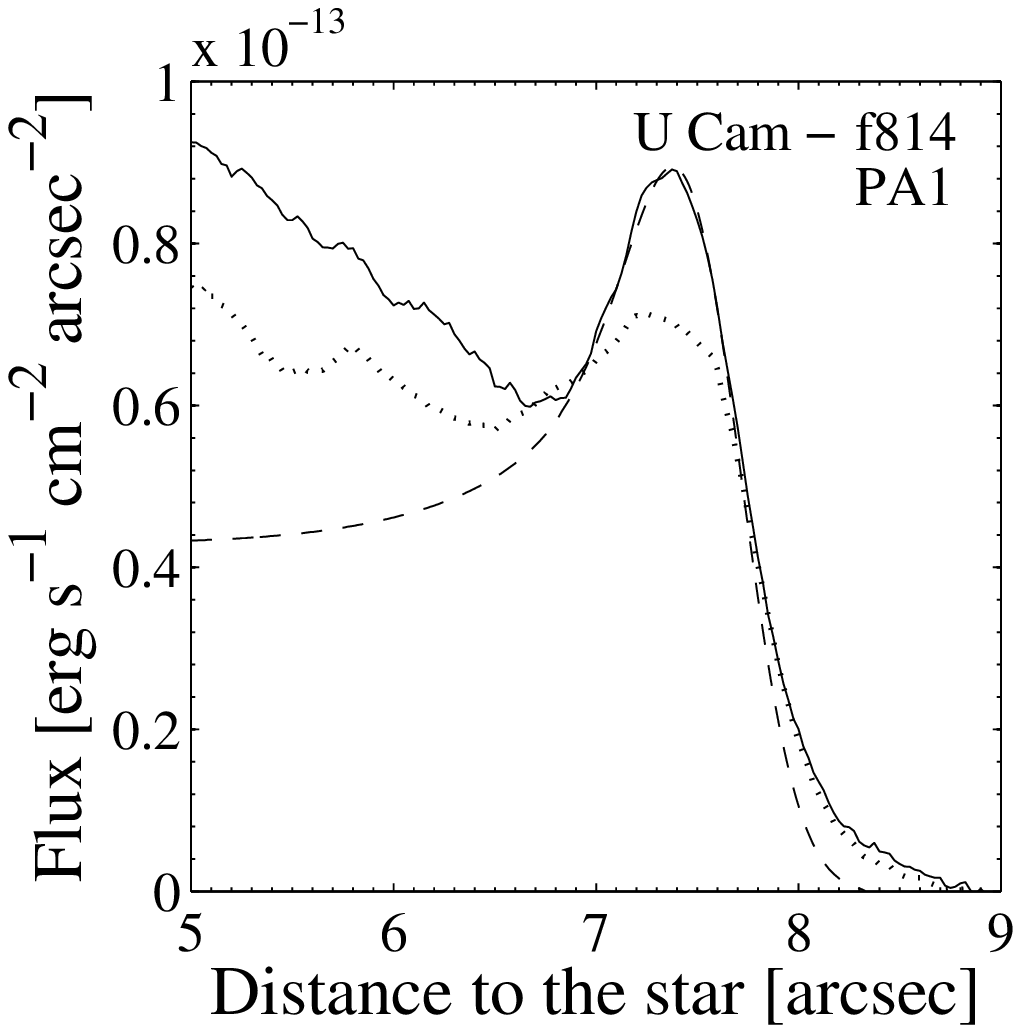}
   \caption{HST/ACS AARPs of U~Cam in the f606 (top) and f814 (bottom) filters in the PA1 interval, the AARPs averaged over all PAs (dotted line), and fits assuming scattering of stellar light in a detached shell of dust (dashed line). See text for details. }
   \label{f:ucamaarps}
\end{figure}

\begin{table}
\caption{Results of the fits to the AARPs in different PA intervals for R~Scl and U~Cam.}
\label{t:resultspa}
\centering
\begin{tabular}{ l c c c c}
\hline\hline
filter 	& PA 	& $R$	& $\Delta R$	& $I_{\rm{peak}}$ \\
		& interval &[\arcsec]	& [\arcsec]		& [erg\,s$^{-1}$\,cm$^{-2}$\,{\AA}$^{-1}$\,$\arcsec^{-2}$]\\
\hline
\multicolumn{2}{l}{R~Scl} & & &\\
f475	& PA2	& 18.5	& 0.9			& 0.7$\,\times\,10^{-17}$\\	
	& PA5	& 19.0	& 0.9			& 0.4$\,\times\,10^{-17}$\\
f606	& PA1	& 18.7	& 1.8			& 6.3$\,\times\,10^{-17}$\\
	& PA2	& 18.9	& 1.6			& 6.2$\,\times\,10^{-17}$\\
	& PA3	& 19.5	& 1.1			& 5.5$\,\times\,10^{-17}$\\
	& PA4	& 20.0	& 0.8			& 5.0$\,\times\,10^{-17}$\\
	& PA5	& 19.3	& 1.6			& 4.8$\,\times\,10^{-17}$\\
f814	& PA1	& 18.8	& 1.8			& 6.0$\,\times\,10^{-17}$\\
	& PA2	& 18.9	& 1.6			& 5.9$\,\times\,10^{-17}$\\
	& PA3	& 19.6	& 1.1			& 5.1$\,\times\,10^{-17}$\\
	& PA4	& 19.9	& 0.7			& 4.8$\,\times\,10^{-17}$\\
	& PA5	& 19.4	& 1.9			& 4.6$\,\times\,10^{-17}$\\
\hline
\multicolumn{2}{l}{U~Cam}& & &\\
f606	& PA1	& 7.6		& 0.5			& 1.3$\,\times\,10^{-17}$ \\
	& PA2	& 7.7		& 0.4			& 1.2$\,\times\,10^{-17}$\\
	& PA3	& 7.7		& 0.7			& 0.9$\,\times\,10^{-17}$\\
f814	& PA1	& 7.6		& 0.6			& 3.6$\,\times\,10^{-17}$\\
	& PA2	& 7.7		& 0.5			& 2.9$\,\times\,10^{-17}$\\
	& PA3	& 7.6		& 0.8			& 2.8$\,\times\,10^{-17}$\\
\hline
\end{tabular}
\end{table}

\subsubsection{U~Cam}
\label{s:ucamappear}

In this case, the ACS in high-resolution mode covers the full extent of the circumstellar
shell (Fig.~\ref{f:ucamimages}). However, the considerably smaller extent means that the artefacts
caused by the central star light covers a larger fraction of the circumstellar
emission. In addition, since observations were obtained at only one rotation angle
of the satellite, the artifact due to the 3$\arcsec$ coronographic spot can be
seen at the north-western edge of the CSE. Nevertheless, a limb brightening is clearly visible and the deviations from a circle of this is less than 10\%, i.e., it is consistent with a CSE that has the form of a geometrically thin shell with overall spherical symmetry.

As for R~Scl we create AARPs of the circumstellar emission around U~Cam over the entire image and in different position angles. In this case the PA intervals had a width of 120$^{\circ}$ and are centred on 60$^{\circ}$ (PA1), 180$^{\circ}$ (PA2), and 300$^{\circ}$ (PA3) (see Fig.~\ref{f:ucamimages}). The region affected by the 3$\arcsec$ coronographic spot is not included in PA3. Figure~\ref{f:ucamaarps} shows the resulting AARPs for PA1 and the AARPs over the entire image in the f606 and f814 filters.

Similarly to the case of R~Scl, the surface brightness in the AARP of the entire image is uniform out to a rather sharp outer edge. The limb brightening, however, is much more apparent here (particularly in the f606 image). 

There is considerable patchiness in the brightness distributions, which is fully
consistent in the f606 and f814 images. There appears to be no large-scale dominant features,
but the outer edge clearly exhibits a kind of undulating pattern, which may
simply be due to a random distribution of varying surface brightness in the shell. 


\subsection{Scattering in detached dust shells}
\label{s:dustscat}

\citet{maeretal10} analysed the  scattered light around the carbon star U~Ant by fitting calculated radial intensity distributions to the observed AARPs using the optically thin approximation for scattering of stellar light by dust grains in geometrically thin circumstellar shells. They assumed that the grains are of a single size (0.1\,$\mu$m), consist of amorphous carbon, and have a density of 2\,g\,cm$^{-3}$. The optical constants were taken from \citet{suh00} $-$ who analysed the spectral energy distributions (SEDs) of carbon stars. The scattering efficiency (as a function of scattering angle) was calculated with the Mie theory. The radial density distribution of the dust wass assumed to follow a Gaussian distribution centred at a radius $R$ with a full width at half maximum $\Delta R$. We follow the same procedure as in \citet{maeretal10} to derive the size and width of the detached shells around R~Scl and U~Cam. Calculated brightness distributions are fitted to the AARPs in the different PA intervals for R~Scl and U~Cam. Examples of the fits are shown in Figs.~\ref{f:rsclaarps} and~\ref{f:ucamaarps}.  Here we adopt a grain size of 0.15\,$\mu$m, because the higher forward scattering efficiency of these grains gives a better fit to the AARPs for both R~Scl and U~Cam (see below).

In general, the peak due to limb brightening and the tail could be fitted reasonably well, giving good estimates of the shell radius and width. The results for $R$ and $\Delta R$ at the different PA intervals for R~Scl and U~Cam are given in Table~\ref{t:resultspa}. On average, the dust shell radii for R~Scl and U~Cam are 19$\farcs$2 (corresponding to a linear radius of 8$\times$10$^{16}$\,cm) and 7$\farcs$7 (corresponding to a linear radius of 5$\times$10$^{16}$\,cm), respectively. The average dust shell widths based on the fits are 1$\farcs$2 (corresponding to a linear width of 5$\times$10$^{15}$\,cm) and 0$\farcs$6 (corresponding to a linear width of 4$\times$10$^{15}$\,cm) for R~Scl and U~Cam, respectively. Modelling of the polarised scattered flux from R~Scl gives a shell radius of 20$\arcsec$, and a width of 2$\arcsec$ \citep{gonzetal03a}. The ground-based observations are subject to seeing, which has the effect of moving the peak to larger radii and widening the shell, and so the results are consistent with each other. \citet{schoetal05b} derive a dust shell radius of $\approx$\,27$\arcsec$ by modelling the SED of R~Scl, but with considerable uncertainty ($\approx$\,$\pm$14$\arcsec$). A recent SMA CO($J$\,=\,2--1) interferometer map of the R~Scl shell gives an estimated CO shell radius of $\approx$\,18$\arcsec$ at a resolution of about 4$\arcsec$ (Dinh-van-Trung, priv. comm.). For U~Cam the size of the CO shell determined in IRAM PdB interferometer maps is 7$\farcs$3, which excellently agrees with the results derived here \citep{lindetal99}. The width is more difficult to derive in the interferometer data, but is estimated to be $\approx$1$\arcsec$. Thus, we conclude that in both cases the dust and gas shells coincide spatially within the uncertainties ($\le$\,1$\arcsec$ and $\approx$\,2$\arcsec$ in the case of R~Scl and U~Cam, respectively). This contrasts with the finding for the detached shell(s) around the carbon star U~Ant, where the gas and dust shells have clearly separated \citep{gonzetal03a, maeretal10}.

\begin{table*}[t]
\caption{Quantitative results for R~Scl and U~Cam from the fits to the observed brightness distributions averaged over all PAs $^{a}$.}
\label{t:results}
\centering
\begin{tabular}{llccccccc}
\hline\hline
Source	& filter 	& $R$ 	& $\Delta R$ 	& $F_\ast $ 	& $I_{\rm CSE}$ &	$F_{\rm CSE}$ & $S_{\rm CSE}$ 	& $F_{\rm CSE}/F_\ast$\\
      		&   		& [\arcsec]	& [\arcsec]     	& [erg\,s$^{-1}$\,cm$^{-2}$]  & [erg\,s$^{-1}$\,cm$^{-2}$\,{\AA}$^{-1}$\,$\arcsec^{-2}$]
               & [erg\,s$^{-1}$\,cm$^{-2}$] & [erg\,s$^{-1}$\,cm$^{-2}$\,{\AA}$^{-1}$] & \\
\hline
R Scl 	& f475 & 18.8 & 0.9 & 4.9\,$\times$\,10$^{-9}$ & 0.5\,$\times$\,10$^{-17}$ & 0.1\,$\times$\,10$^{-10}$&0.6\,$\times$\,10$^{-14}$ & 17.0\,$\times$\,10$^{-4}$ \\
     		& f606 & 19.3 &1.4 & 1.3\,$\times$\,10$^{-7}$ & 4.9\,$\times$\,10$^{-17}$ & 1.4\,$\times$\,10$^{-10}$ &6.0\,$\times$\,10$^{-14}$ & 11.1\,$\times$\,10$^{-4}$ \\
     		& f814 & 19.3 & 1.4 & 1.3\,$\times$\,10$^{-7}$ & 4.4\,$\times$\,10$^{-17}$ & 	1.4\,$\times$\,10$^{-10}$ &5.5\,$\times$\,10$^{-14}$ &10.3\,$\times$\,10$^{-4}$ \\
U Cam 	& f606 &\phantom{1}7.7 & 0.5 &  1.4\,$\times$\,10$^{-8}$ & 0.6\,$\times$\,10$^{-17}$ & 	0.4\,$\times$\,10$^{-11}$	&0.2\,$\times$\,10$^{-14}$ & \phantom{1}2.6\,$\times$\,10$^{-4}$ \\
     		& f814 &\phantom{1}7.6 & 0.6 & 5.4\,$\times$\,10$^{-8}$ & 3.0\,$\times$\,10$^{-17}$ & 1.4\,$\times$\,10$^{-11}$		&0.6\,$\times$\,10$^{-14}$ & \phantom{1}2.5\,$\times$\,10$^{-4}$\\
\hline
\end{tabular}
\begin{list}{}{}
\item[$^{\mathrm{a}}$] The columns give the radius ($R$) and the FWHM ($\Delta R$) of the Gaussian dust density distribution, the stellar flux ($F_{\ast}$), the average surface brightness ($I_{\rm{CSE}}$), the flux of the scattered stellar light ($F_{\rm{CSE}}$), the flux density of the scattered stellar light ($S_{\rm{CSE}}$), and the ratio between the circumstellar and stellar fluxes ($F_{\rm{CSE}}/F_{\ast}$). See text for details.
\end{list}
\end{table*}

The peak brightness of the limb brightening is given for the different PA intervals in Table~\ref{t:resultspa}. It is clear that the limb brightening is much less pronounced in the AARP averaged over all PAs for both R~Scl and U~Cam. For R~Scl the fits to the AARPs in the different PA intervals show that the estimated radius of the shell varies by $\approx$\,$\pm$1$\arcsec$ (caused by the centre of the shell most likely not being centred at R~Scl). At the same time the width of the shell is $\approx$\,1$\farcs$2, with the consequence that the limb brightening in the AARP over the entire image is smeared out. The estimated radius of the shell around U~Cam is essentially the same in all PA intervals, but the width varies considerably, which also leads to a smearing of the limb brightening in the AARP averaged over all PAs.

The scattering of the grains occurs preferentially in the forward direction when the grain size becomes comparable to the observed wavelength [the details of the scattering and its angular dependence is described in \citet{maeretal10}], and this explains that the scattered light does not become significantly weaker closer to the star which would otherwise be expected for isotropic scattering in a geometrically thin shell. Indeed, this property can be used to estimate the size of the grains because their forward scattering efficiency depends on this (as is shown in Fig.~\ref{qscat} through $g$, which is a measure of the directional dependence of the scattering). However, it is clear that the simplified model does not fully explain the AARPs at line-of-sights closer to the star. We find a reasonable fit to the AARPs for a grain size of 0.15\,$\mu$m, but there is a notable systematic underestimate in the f814 filter. A distribution of grain sizes would probably lead to a better fit in all filters. Also, the details of the AARPs are not well fit. This is likely due to the patchiness observed in the shells (i.e., the detailed density distribution of the dust). The average surface brightnesses of the shells in the different filters are therefore obtained as straight averages over the area outside a radius of 5$\arcsec$ and inside a radius of $\approx$\,17$\arcsec$ and 7$\arcsec$ for R~Scl and U~Cam, respectively. The resulting values are given in Tables~\ref{t:resultspa} (for the different PA intervals) and~\ref{t:results} (for the entire image). For R~Scl, \citet{gonzetal01} obtained 2\,$\times$\,10$^{-17}$  and 4\,$\times$\,10$^{-17}$ (in units of erg\,s$^{-1}$\,cm$^{-2}$\,{\AA}$^{-1}$\,$\arcsec^{-2}$; with some considerable uncertainty) in 50{\AA} wide filters centred at 590\,nm and 770\,nm, respectively (i.e., the wavelength ranges of these filters lie inside those of the f606 and f814 filters, respectively). Thus, within the uncertainty, the flux densities are the same in the narrow and the broad filters, suggesting that even in the narrow filters the contribution from atomic scattering (by Na and K in the 590\,nm and 770\,nm filter, respectively) is limited, and the broad filters used on the HST are completely dominated by dust scattering.

Of a more quantitative importance is the ratio between the integrated surface brightness and the stellar flux, i.e., the fraction of the stellar light that is scattered. The integrated scattered light fluxes are obtained by summing the AARPs covering the entire image over all angles. The estimated uncertainty in the stellar fluxes due to the psf fitting is estimated to be a factor of $\approx2$. The circumstellar and stellar fluxes as well as the resulting circumstellar to stellar flux ratios are given in Table~\ref{t:results}. The results show that on average the scattering process is very optically thin.


\subsection{Dust shell masses}
\label{s:dustmass}

We proceed with a simple analysis of the observed scattered
emission based on the assumption of optically thin dust scattering in a
geometrically thin shell following the work of \citet{gonzetal01}. The circumstellar scattered flux (at the
distance of the object), integrated over the total scattering wavelength range (defined by filter f) in the
optically thin regime is given by
\begin{equation}
         F_\mathrm{sc,f}
         = \frac{1}{4 \pi R_\mathrm{sh}^2} N_\mathrm{sc} \int_{\rm f} \sigma_{\mathrm{sc},\lambda}
         F_{\ast \lambda} {\rm d}\lambda\,,
\end{equation}
where $F_{\ast \lambda}$ is the stellar surface flux. Hence, the expected circumstellar to stellar flux ratio is given by
\begin{equation}
\label{e:cssthin}	 
        \frac{F_{\mathrm{sc,f}}}{F_{\ast,{\rm f}}} = 
	          \frac{1}{4\pi R_\mathrm{sh}^2}
              N_\mathrm{sc} \frac{\int_{\rm f} \sigma_{\mathrm{sc},\lambda}
              F_{\ast \lambda} {\rm d}\lambda}{F_{\ast,{\rm f}}}\,,
\end{equation}
where $N_\mathrm{sc}$ is the number of scatterers, $\sigma_\mathrm{sc}$
the scattering cross section, and $R_\mathrm{sh}$ the shell
radius.

For dust scattering, and assuming single-sized,
spherical grains, we have the following relations for the number of
scatterers and the scattering cross section
\begin{equation}
     \label{e:dustscatterers}	 
        N_\mathrm{sc}=\frac{3 M_\mathrm{d}}{4\pi \rho_\mathrm{g} a^3}\,,
\end{equation}
\begin{equation}
     \label{e:dustcrossection}	 
        \sigma_{\mathrm{sc},\lambda} = Q_{\mathrm{sc},\lambda} \pi a^2\,,
\end{equation}
where $M_\mathrm{d}$ is the dust mass in the shell, $\rho_\mathrm{g}$
the density of a dust grain, $a$ the grain radius, and $Q_{\mathrm{sc},\lambda}$ 
the grain scattering efficiency (assuming isotropic scattering). $Q_{\rm{sc,\lambda}}$ is calculated with the Mie theory and the optical constants from \citet{suh00} for carbonaceous dust grains with a radius of $0.15\,\mu$m and a grain density of $2\,\rm{g\,cm^{-3}}$.

\begin{figure}
\centering
      \includegraphics[width=7cm]{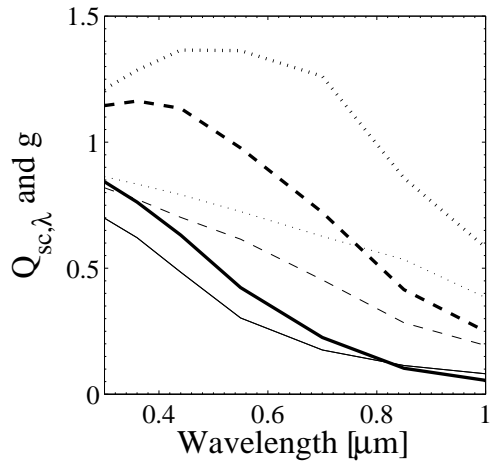}
\caption{The scattering efficiency $Q_{\rm{sc,\lambda}}$ (thick lines) and a measure of the directional dependence of the scattering $g$ (thin lines) as functions of wavelength for 0.1$\,\mu$, 0.15$\,\mu$m, and 0.2$\,\mu$m sized grains (solid, dashed, and dotted  lines, respectively).}
\label{qscat}
\end{figure}

$Q_{\rm{sc,\lambda}}$ varies strongly with wavelength and grain size, see Fig.~\ref{qscat}. Consequently, $Q_{\rm{sc,\lambda}}$ varies considerably over the wavelength range covered by one filter. A directional dependence of the scattering, which depends on the grain size, further complicates the problem. As opposed to the analysis of the AARPs we here assume isotropic scattering for reasons of simplicity. Finally, the assumption of one constant grain size throughout the shell adds an additional uncertainty. This results in significant uncertainties in the determined dust masses in the shell. Together with the uncertainty in the stellar fluxes (and hence $F_{\rm CSE}/F_*$ ratios), we estimate the resulting dust masses to be uncertain by a factor $\approx$\,5. The resulting shell masses based on the measured circumstellar to stellar fluxes for R~Scl and U~Cam are 3$\times$10$^{-6}$\,$M_{\odot}$ and 3$\times$10$^{-7}$\,$M_{\odot}$, respectively (Table~\ref{t:masses}). For R~Scl, \citet{gonzetal03a} derived a dust mass, based on models of polarised scattered light, of 2$\times$10$^{-6}$\,$M_{\odot}$, while \citet{schoetal05b} derived a dust mass of (3.2\,$\pm$\,2.0)$\times$10$^{-5}$\,$M_{\odot}$ by modelling the dust thermal emission. For U~Cam \citet{schoetal05b} derived only an upper limit to the dust mass. \citet{schoetal05b} derived gas shell masses, based on CO radio line modelling, of 2.5$\times$10$^{-3}$\,$M_{\odot}$ and 1$\times$10$^{-3}$\,$M_{\odot}$ for R~Scl and U~Cam, respectively, leading to dust-to-gas mass ratios of 1$\times$10$^{-3}$ and 3$\times$10$^{-4}$, rather low values for carbon stars \citep{ramsetal08a}.

\subsection{The small-scale structure of the circumstellar medium}
\label{patches}

The detached, geometrically thin shells observed here are excellent for studies of the detailed density distribution of the circumstellar medium, in particular in dust-scattered stellar light, where the optical depth is low. For both R~Scl and U~Cam significant patchiness is observed in the images, and it is close to identical in the different filters for the respective object. Due to the smaller size of the shell the patchiness is less obvious for of U~Cam. In particular the artefacts due to the psf-subtraction in the inner parts of the shell make it difficult to analyse the U~Cam data.  

\begin{figure*}[t]
   \centering
   \includegraphics[width=8cm]{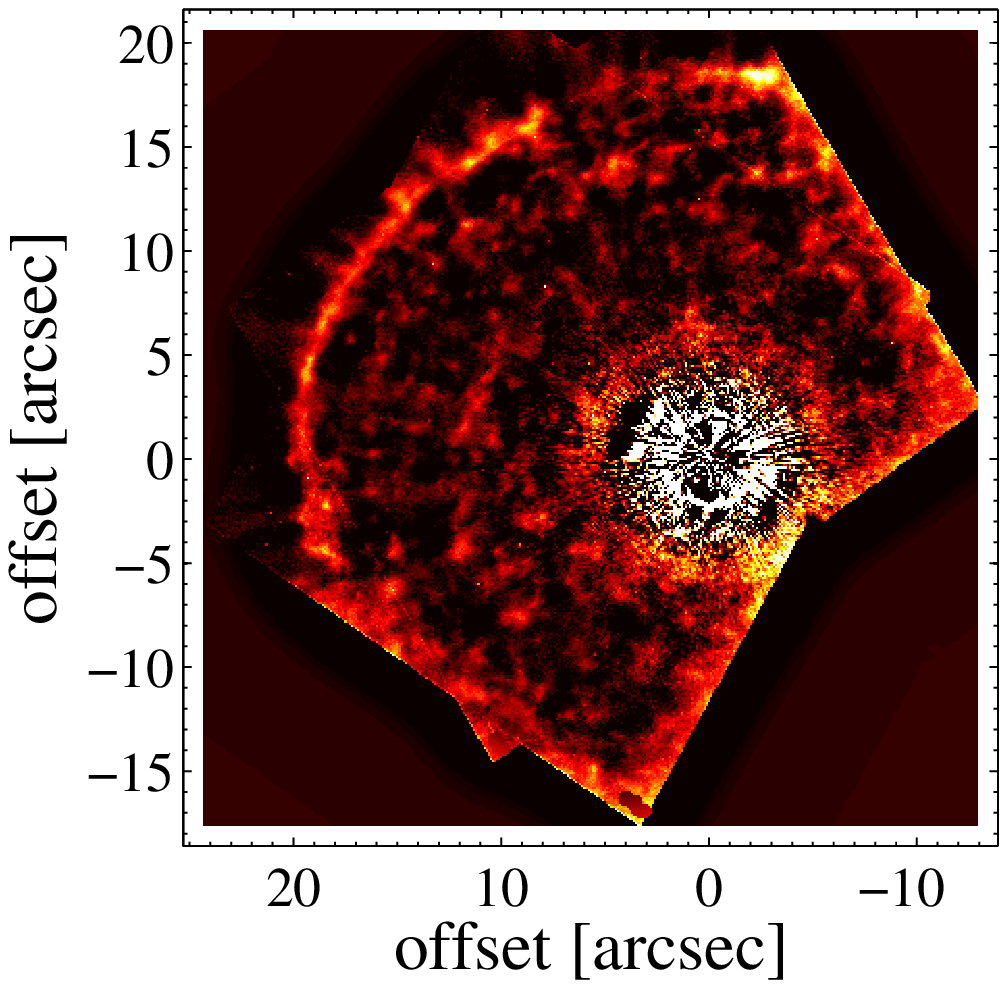}
   \includegraphics[width=8cm]{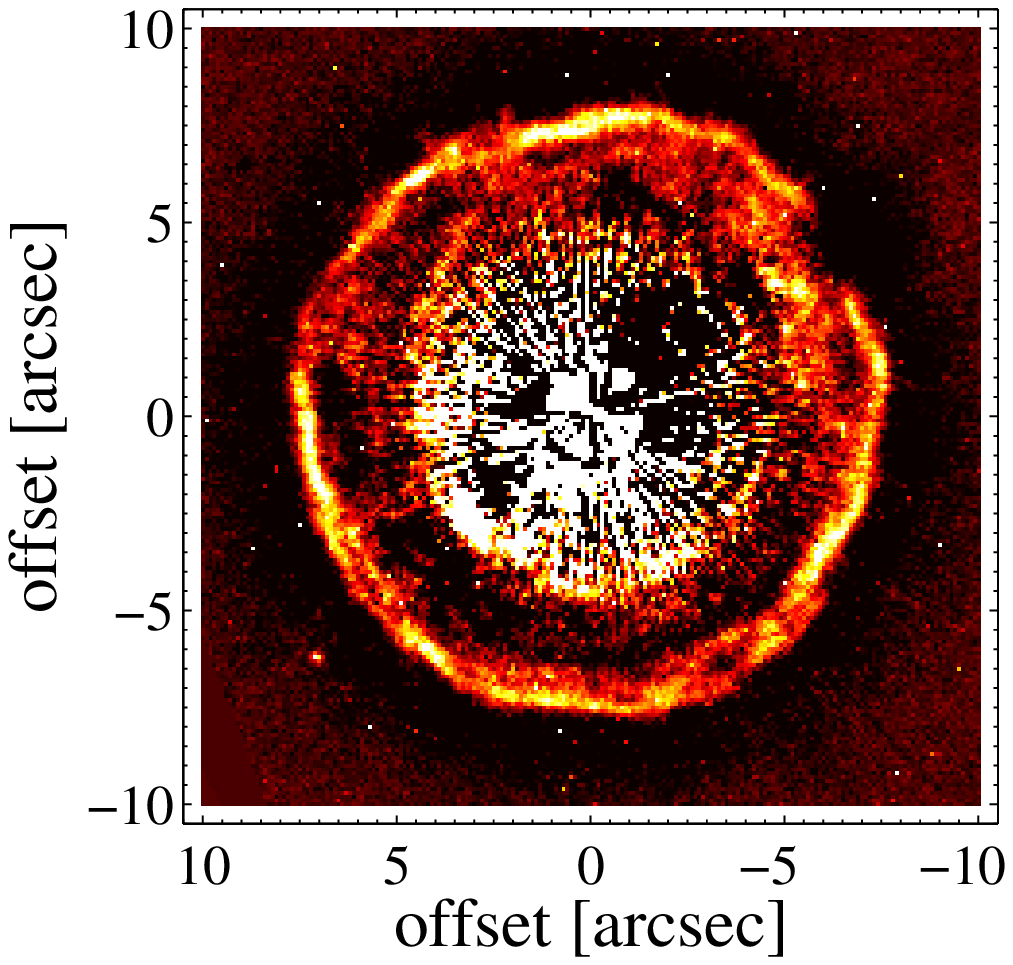} 
   \caption{Images in the f814 filter of R~Scl (right) and U~Cam (left) after the subtraction of an unsharp mask, emphasising the structure in the detached shells.}
   \label{f:sharp}
\end{figure*}

\begin{figure}
\centering
      \includegraphics[width=7cm]{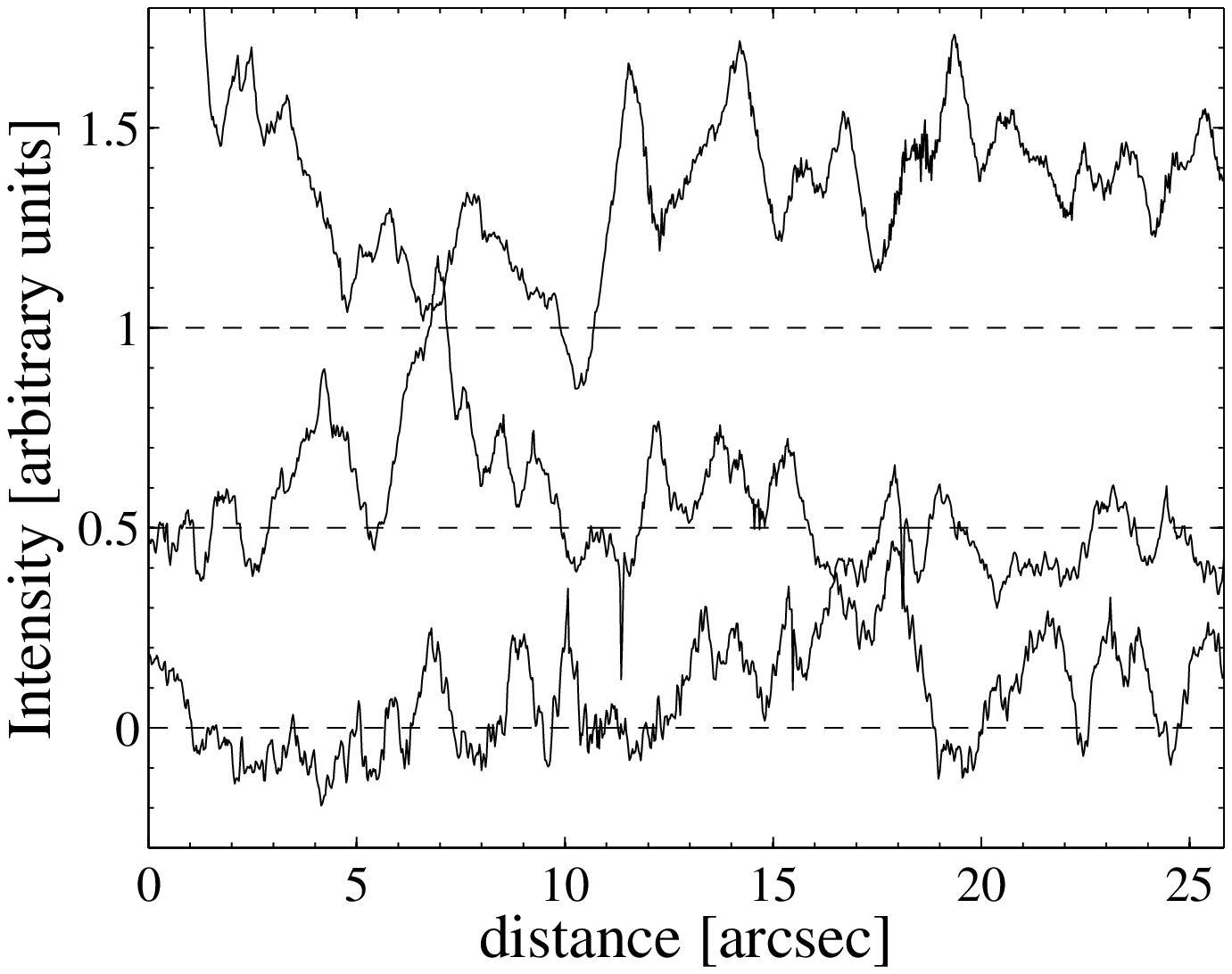}
      \includegraphics[width=7cm]{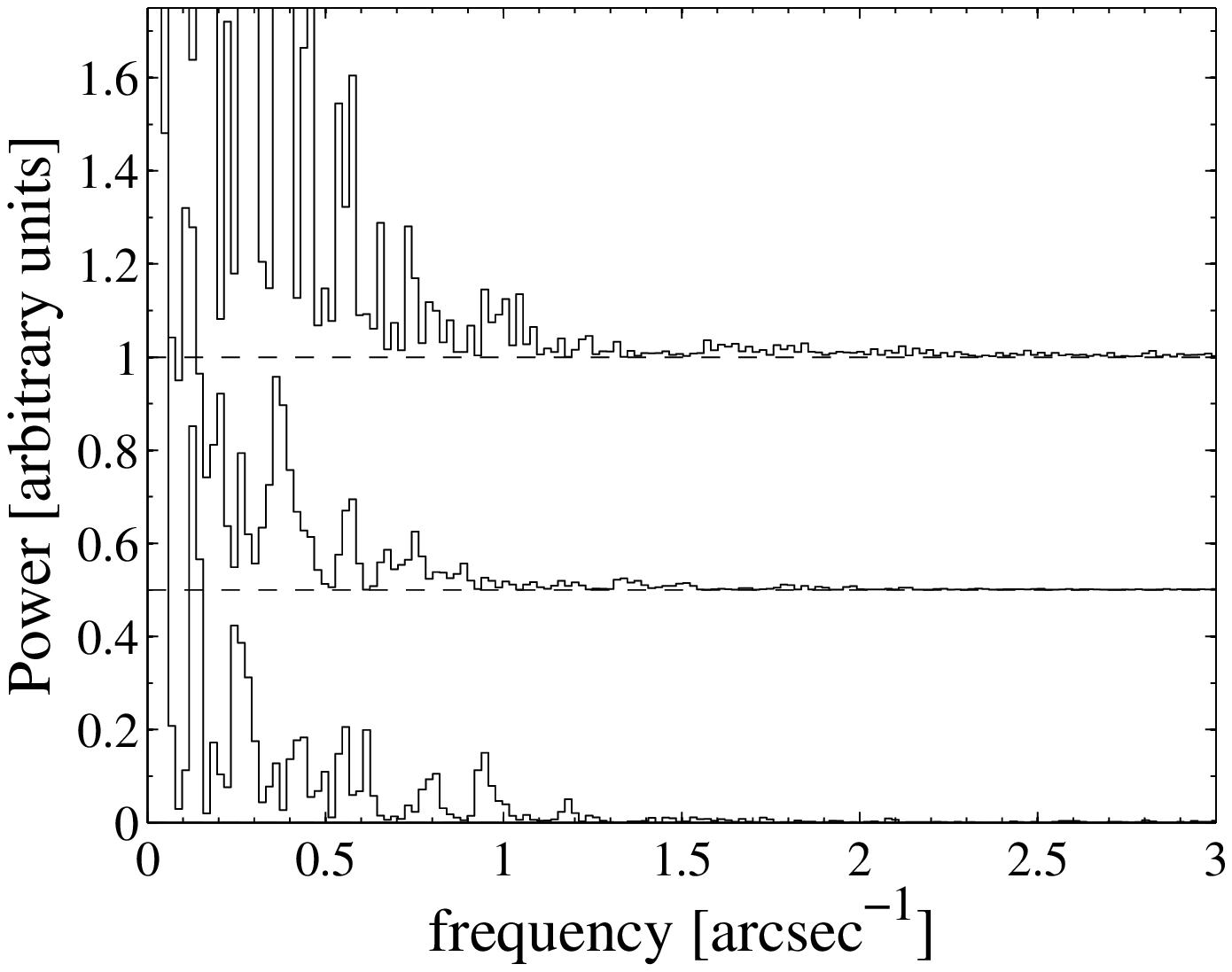}
\caption{\emph{Top panel:} RAAPs at three distances from R~Scl: 19\farcs1 (top), 15\farcs5 (middle), and 11\farcs5 (bottom). The azimuthal profiles are averaged over a radial range of 5\% of the radius. \emph{Bottom panel:} Corresponding power spectra for the different profiles (in the same order from top to bottom as in the upper panel). The profiles in both panels are offset for clarity. The original zero levels are indicated by the dashed lines.}
\label{f:rsclraaps}
\end{figure}

\begin{figure}
\centering
      \includegraphics[width=7cm]{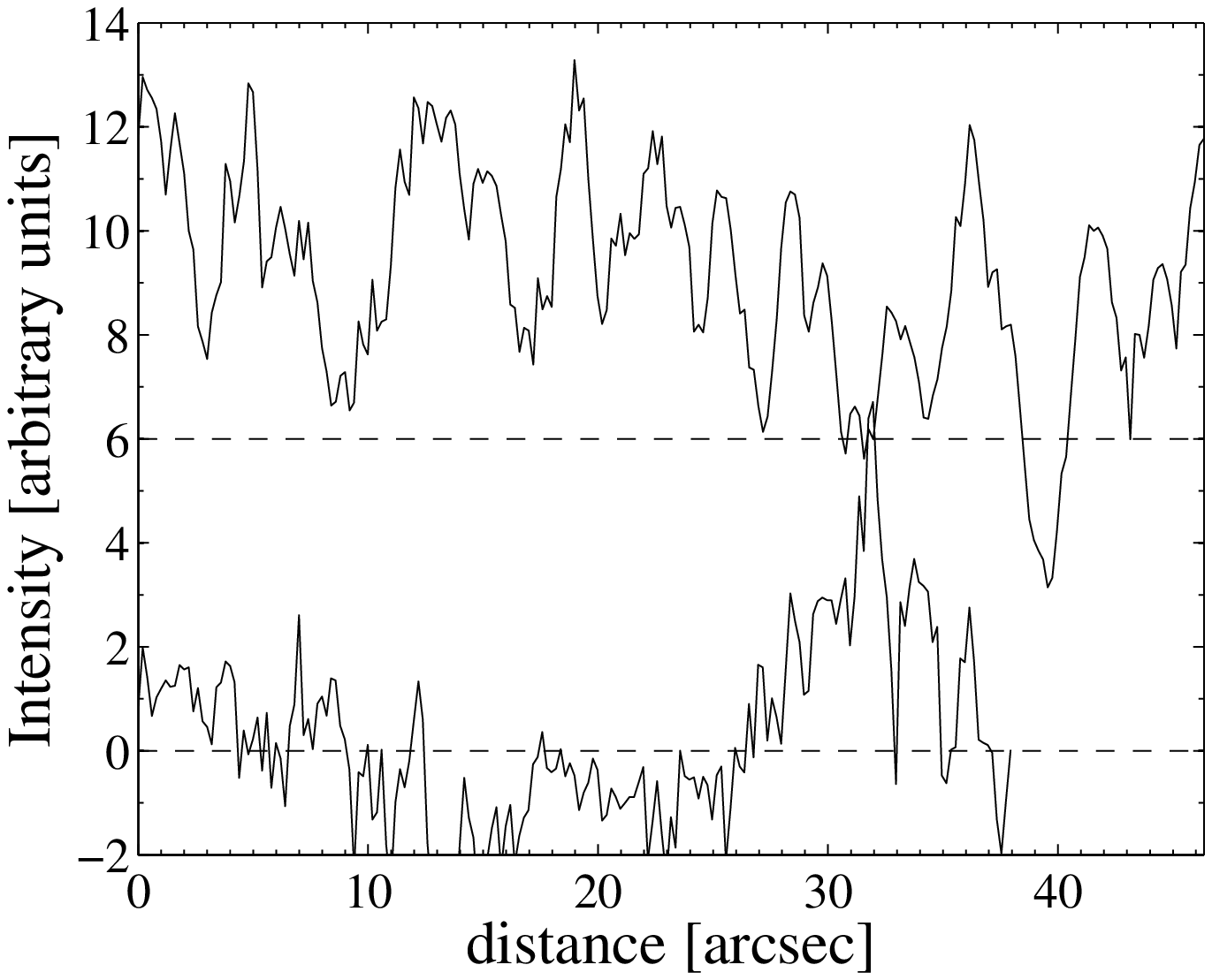}
      \includegraphics[width=7cm]{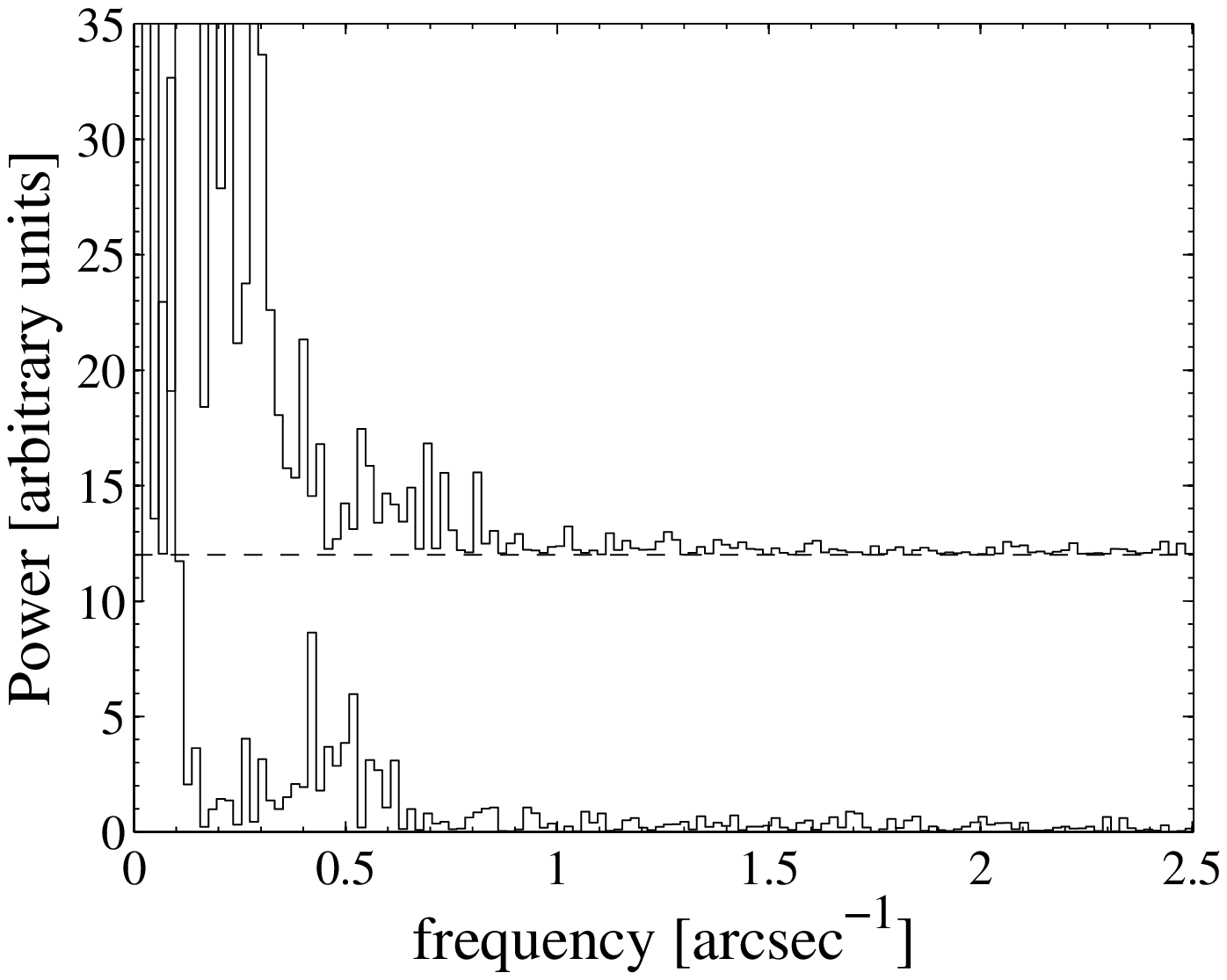}
\caption{\emph{Top panel:} RAAPs at two distances from U~Cam: 7\farcs4 (top) and 6\farcs1 (bottom). The azimuthal profiles are averaged over a radial range of 5\% of the radius. \emph{Bottom panel:} Corresponding power spectra for the different profiles (in the same order from top to bottom as in the upper panel). The profiles in both panels are offset for clarity. The original zero levels are indicated by the dashed lines.}
\label{f:ucamraaps}
\end{figure}

In order to analyse the small-scale structure in more detail, unsharp masks (smoothing with a Gaussian having a FWHM of 3.5$\arcsec$) were subtracted from the f814 images for R~Scl and U~Cam, Fig.~\ref{f:sharp}. In these images the ``clumpiness'' is much more apparent.
Radially averaged azimuthal profiles (RAAPs) were created from the sharp images at three radial distances for R~Scl (11\farcs5, 15\farcs5, and 19\farcs1) and two radial distances for U~Cam (6\farcs1 and 7\farcs4). The azimuthal ranges covered by the RAAPs for R~Scl and U~Cam are indicated in Figs.~\ref{f:rsclimages} and~\ref{f:ucamimages}, respectively. The RAAPs were averaged over a radial range corresponding to 5\% of the radial distance. The resolution along the azimuthal direction is 0\farcs025 for R~Scl and 0\farcs2 for U~Cam. The resulting RAAPs are shown in the upper panels of Figs.~\ref{f:rsclraaps} and~\ref{f:ucamraaps} for R~Scl and U~Cam, respectively. The patchy structure can clearly be seen in the RAAPs of R~Scl, and the profiles at the different radii show the same general type of structure, indicating that the clumps are evenly distributed over the shell. The U~Cam image is noisier than the R~Scl image, and the artifacts from the coronograph cover a large fraction of the shell, but the patchy structure can clearly be seen also here in the RAAP at 7\farcs4. The lower panels show the corresponding Fourier transforms of the RAAPs. They show that some tendencies for structures arise on scales $\ge$\,1$\arcsec$ in both cases.

\begin{table}
\caption{The derived dust shell-masses based on the observed $F_{\rm CSE}/F_*$ ratios.}
\label{t:masses}
\centering
\begin{tabular}{llcc} 
\hline\hline
Source 	& filter 	& $M_{\rm{d}}$ &$M_{\rm{g}}\,^a$\\
		&		&Ê[$10^{-6}\,M_{\odot}$] & [$10^{-3}\,M_{\odot}$]\\
\hline
R~ÊScl 	& f475 	& 2.6 &2.5\\
		& f606	& 2.1 &\\
		& f814	& 4.0 &\\
U~Cam	& f606	& 0.2Ê&0.9\\
		& f814	& 0.3 &\\
\hline
\end{tabular}
\begin{list}{}{}
\item[$^{\mathrm{a}}$] The shell gas mass estimated from CO radio line modelling \citep{schoetal05b}. 
\end{list}
\end{table}

\begin{figure}
   \centering
   \includegraphics[width=8.5cm]{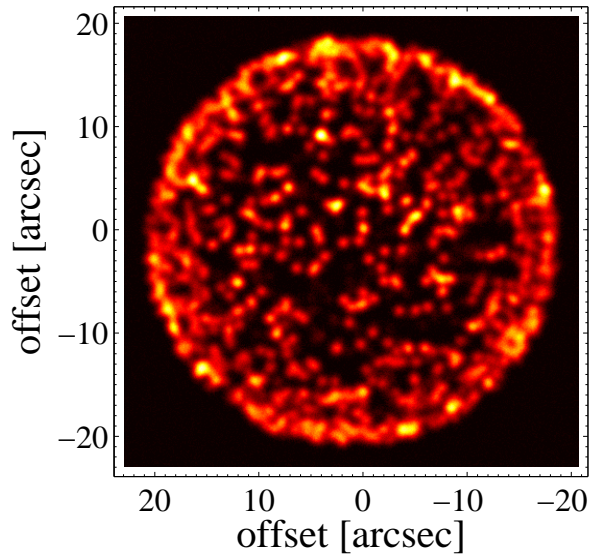}
   \caption{A simulated image of the detached shell around R~Scl seen in dust-scattered light in the f814 filter (see the text for details).}
   \label{f:simimage}
\end{figure}

\begin{figure}
   \centering
   \includegraphics[width=7cm]{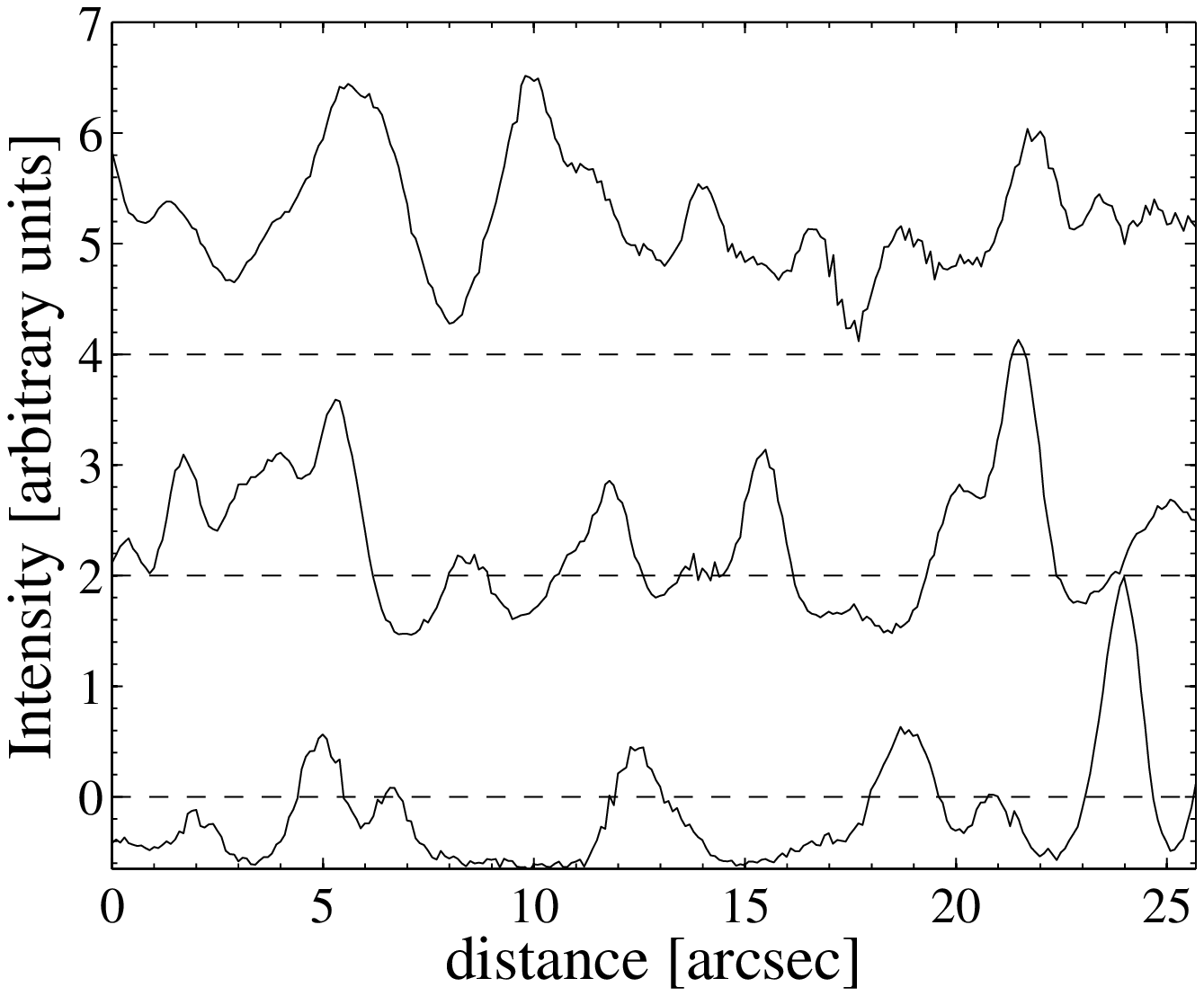}
   \includegraphics[width=7cm]{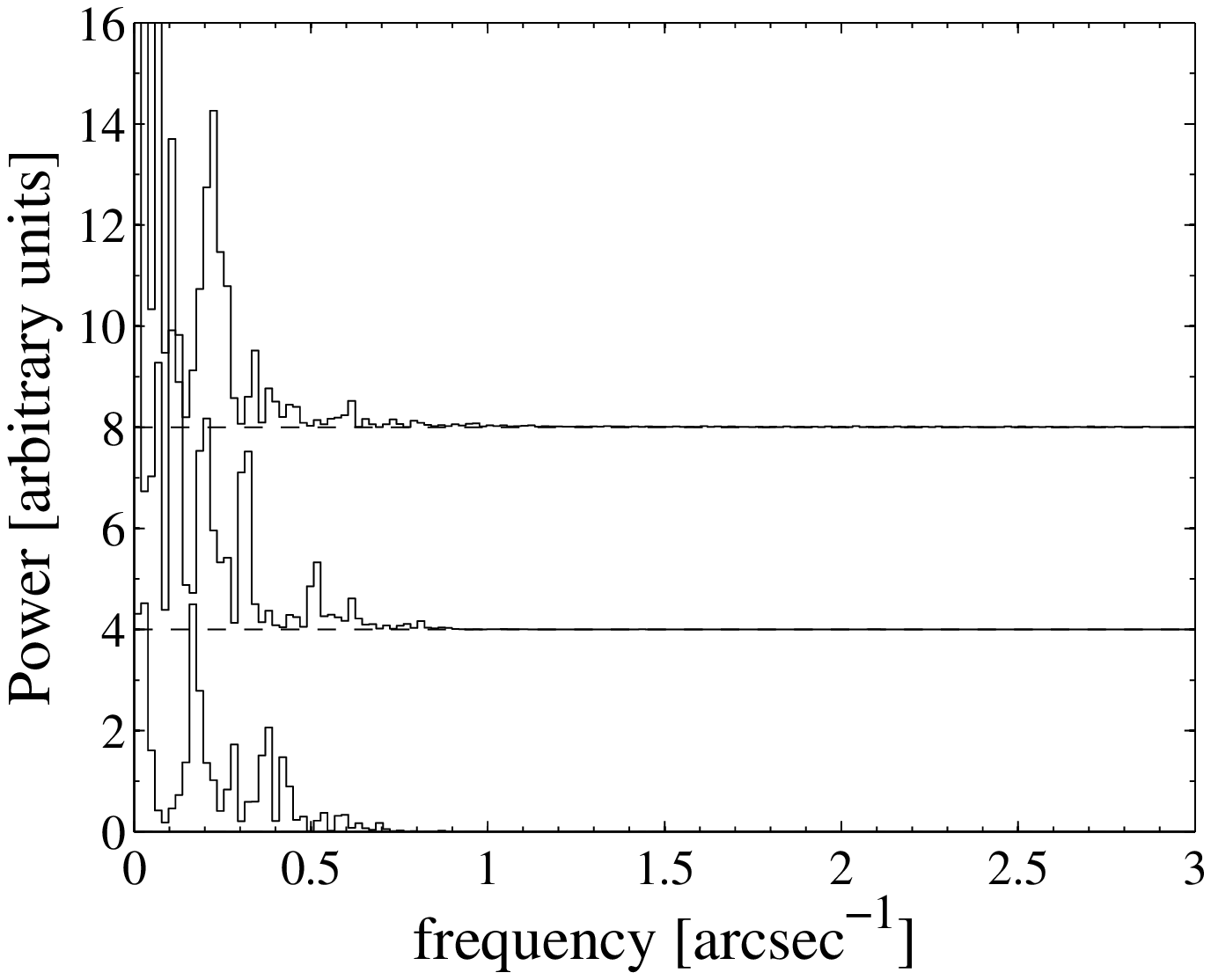}
   \caption{Same as Fig.~\ref{f:rsclraaps}, but with the data input from the sharpened version of the simulated image in Fig.~\ref{f:simimage} (see text for details).}
   \label{f:simraaps}
\end{figure}

The patches seen in the images are here interpreted as signs of a small-scale structure in the form of clumps in the detached shells. We made a crude estimate of the "clumpiness" in the R~Scl shell through randomly distributing clumps of an equal size in a shell of a radius 19$\farcs$2 and a width 1$\farcs$2. Each clump produces optically thin scattering with a directional dependence according to the grain properties used in Sect.~\ref{s:dustscat}. A reasonable resemblance with the observed images and AARPs is found for 2000 clumps of a size 0$\farcs$9 (the FWHM of a Gaussian clump). Structures on scales $\ge$\,1$\arcsec$ arise because of overlapping clumps. As an example, the simulated image in the f814 filter is shown in Fig.~\ref{f:simimage}. The RAAPs, in the same radial and azimuthal ranges used for Fig.~\ref{f:rsclraaps}, obtained from the sharpened version of Fig.~\ref{f:simimage} (using the same procedure as for the observed image) are shown in Fig~\ref{f:simraaps} together with the associated frequency distribution. These are also similar to the observed ones except for a lack of structure on smaller scales, probably an effect of a distribution of clump sizes. \citet{bergetal93} and \citet{olofetal00} estimated that the detached CO shells of the carbon stars S~Sct and TT~Cyg are composed of about 1000 clumps of a size of about 0$\farcs$5 (FWHM), in line with the findings here. Adopting a value of 2000 clumps leads to an estimated clump dust mass of about 2$\times$10$^{-9}$\,$M_\odot$ for R~Scl.

A comparison of Figs~\ref{f:rsclimages}, \ref{f:ucamimages}, and \ref{f:sharp} with Fig.~\ref{f:simimage} shows a difference between the simulations and the observations. The observations show sections of possible double-shell structures or features, which appear to be fragmented shells or possibly filaments, as opposed to the simulations where the clumps are randomly distributed in the shell. It is unclear if these structures are the results of wind-wind interactions due to instabilities, or of density or velocity inhomogeneities in any of the two winds, or both.

\section{Discussion and conclusions}

\subsection{Characteristics of the detached shells}

At the distances of R~Scl and U~Cam the angular radii of the shells correspond to linear radii of 8$\times$10$^{16}$\,cm and 5$\times$10$^{16}$\,cm, respectively. \citet{schoetal05b} estimate that the expansion velocities for the gas shells of R~Scl and U~Cam are 15.5\,km\,s$^{-1}$ and 23.0\,km\,s$^{-1}$, respectively, and, since there appears to be no separation between the gas and the dust shells, the corresponding dynamical ages of the shells are consequently about 1700 yr (R~Scl) and 700 yr (U~Cam). Thus, these are the youngest detached shells observed in detail so far. The average shell widths convert into linear widths of 5$\times$10$^{15}$\,cm and 4$\times$10$^{15}$\,cm for R~Scl and U~Cam, respectively, and the corresponding time scales are 100 yr (R~Scl) and 50 yr (U~Cam). The shell width/radius ratios lie in the range 0.05\,$-$\,0.1. The estimated dust and gas shell masses are 3$\times$10$^{-6}$\,$M_{\odot}$ and 2.5$\times$10$^{-3}$\,$M_{\odot}$, respectively, for R~Scl and 3$\times$10$^{-7}$\,$M_{\odot}$ and 1$\times$10$^{-3}$\,$M_{\odot}$, respectively, for U~Cam. To this we can add the recent result by \citet{maeretal10} for the carbon star U~Ant, where the shell has a dynamical age of 2700 yr, a width corresponding to 140 yr, and the dust and gas masses are estimated to be 5$\times$10$^{-5}$\,$M_{\odot}$ and 2.5$\times$10$^{-3}$\,$M_{\odot}$. Our new dust mass estimates agree with the finding by \citet{schoetal05b} that the shell masses increase with the age of the shell, which is possibly an effect of interaction between the shell and a previous slower stellar wind. The dust-to-gas ratio also increases with the age of the shell (3$\times$10$^{-4}$, 1$\times$10$^{-3}$, and 2$\times$10$^{-2}$ for U~Cam, R~Scl, and U~Ant, respectively). This is curious, because one expects the dust to gradually leave the gas shell as it expands \citep{maeretal10}, at least if the wind is smooth. Clumpiness, possibly arising as the shell expands, and/or wind-wind interaction may be an explanation here as this will affect the gas-grain drift and may lead to accumulation of material.

\subsection{Origin of the detached shells}

It has been argued by various authors that the geometrically thin, detached shells found around carbon stars are the effects of mass-loss rate modulations during thermal pulses \citep{olofetal90, olofetal96, schretal98, gonzetal03a, schoetal05b, mattetal07, maeretal10}. In particular, the detailed models of \citet{mattetal07} show that geometrically thin shells appear under certain circumstances as an effect of a mass-loss eruption and a subsequent interaction with a previous slower wind. The results on R~Scl and U~Cam presented here are consistent with such a scenario. They add further support to the presence of highly isotropic mass loss during the shell formation event, and, if interaction plays a major role, this must apply also to the mass loss prior to the eruption. If this interpretation is correct, both stars must be in the aftermath of a very recent thermal pulse ($<$\,2000 years ago). The shell formation time scale is as short as $<$\,100\,yr. This is an estimate that is likely not affected by interaction with a previous slow wind because of the relative youth of the shells. The amount of mass ejected is relatively small, $\approx$\,10$^{-3}$\,$M_\odot$.

An estimate of the mass-loss rate during the formation of the shells is obtained using the shell gas mass estimates of \citet{schoetal05b} and the time scales estimated here. The results are 3$\times$10$^{-5}$ and 2$\times$10$^{-5}$\,$M_{\odot}$\,yr$^{-1}$ for R~Scl and U~Cam, respectively. This ignores any effects of e.g sweeping up of material. An estimate of the mass-loss rate prior to the shell ejection is difficult to give. We limit ourselves here to an estimate of the density contrast between the shell and the surrounding medium. In the light of optically thin scattering, the density contrast between the shell and the medium outside the shell is at least a factor of 30 in the case of R~Scl (for which the data has the highest S/N). This should be a rough estimate of the mass-loss rate contrast as well. The mass-loss rate following the ejection of the shell is difficult to estimate from our data, because it depends critically on the psf subtraction and the properties of the grain scattering (in particular its angular dependence). \citet{schoetal05b} estimated, based on CO radio line modelling, that the present mass-loss rates are 3$\times$10$^{-7}$ and 2$\times$10$^{-7}$\,$M_{\odot}$\,yr$^{-1}$ for R~Scl and U~Cam, respectively. In conclusion, the mass-loss rate modulations are substantial, of the order of two magnitudes.

\subsection{Origin of clumps in the detached shells}

The small-scale structure of the circumstellar medium is of interest for various reasons, e.g., it will affect the radiative transfer and hence the estimates (like those of mass-loss rates and molecular abundances) obtained from modelling molecular line emission, it will affect the circumstellar chemistry (incl. the photodissociation), and it will affect the efficiency of the mass loss. 

The observational evidence for clumpiness in AGB CSEs is scattered and not very conclusive. High angular resolution near-IR observations certainly suggest a very inhomogeneous medium close to the star, where even proper motions of clumps have been measured \citep{weigetal02}. Far-infrared and radio observations of the circumstellar medium normally do not have the resolution to identify any small-scale structure.  However, observations of in particular SiO and H$_2$O maser emission suggest the presence of clumps also in the CSEs. The very long baseline interferometry (VLBI) observations of SiO
masers towards AGB stars suggest spot sizes as small as a few 10$^{12}$\,cm in regions close to the star \citep{coloetal92}.  Indeed, for reasonable mass-loss rates the
gas within $10^{14}$\,cm of the star must be substantially clumped to
provide the densities required for the SiO masers (about 10$^{13}$\,cm$^{-3}$).  Also H$_2$O masers show spot sizes as small as a few 10$^{12}$\,cm \citep{imaietal97}.  The characteristic OH maser
spot size appears to be a few $10^{14}$\,cm at a radius of a few
$10^{15}$\,cm, i.e., further out in the envelope \citep{chapetal94}. Note though that maser observations are difficult to interpret in terms of a detailed density structure, and they tend to enhance any density contrasts present. Larger-scale patchiness in CO radio line brightness maps of detached shells have been interpreted as a consequence of a large number of small clumps \citep{bergetal93, olofetal96, olofetal00}.

On the other hand, high resolution optical observations of planetary nebulae (PNe) show conclusive evidence of small-scale structure, in particular, the cometary globules seen in the Helix nebula. These are about 1$\arcsec$ in size and are clearly affected by the strong radiation field from the central white dwarf 
\citep{meabetal96, meabetal98, odelhand96}. \citet{huggetal92, huggetal02} detected CO radio line emission from these clumps and estimate their masses to be about 10$^{-5}$\,$M_{\odot}$. It is likely that similar structures are present also in other PNe \citep{wils50, zans55, specetal03}; the Helix happens to be one of the most nearby PNe. \citet{huggmaur02} studied the small-scale structure in two well-known CSEs, those around the carbon star IRC+10216 and the young PN NGC 7027, with the aim to identify at which point the small-scale structure arises. They found no evidence of clumpiness in any of the two envelopes. We note in passing that clumps have also been observed in expanding shells around stars in other circumstances. Examples are O associations \citep{frie54}, HII regions \citep{reipetal97}, and in the environments of young stellar objects \citep{rudowelc88, ketoho89}.

The combination of optically thin scattering due to dust and the narrow widths of the detached shells is excellent for studying the small-scale structure. A potential drawback is that the mass-dominant component, the gas, may not be distributed in the same way as the dust \citep[see e.g.,][]{maeretal10}. The R~Scl sharpened image lends itself to a quantitative analysis of the clump properties. It suggests clump sizes of about 0$\farcs$9 in diameter, corresponding to linear clump diameters of 4$\times$10$^{15}$\,cm. There is an estimated total of about 2000 clumps in the shell. The average clump dust mass is consequently about 2$\times$10$^{-9}$\,$M_{\odot}$. The estimated dust-to-gas ratio suggests a clump gas mass of about 2$\times$10$^{-6}$\,$M_{\odot}$. This is somewhat lower than the mass estimates of the Helix nebula cometary globules \citep{huggetal02}, but the uncertainties in both estimates are substantial.

A likely explanation for the creation of detached shells around carbon stars is an increased mass-loss during the He-shell flash, with a subsequent interaction of a faster wind running into a previous, slower wind \citep{stefscho00, schoetal05b, mattetal07}. The question arises whether the clumpy structure is already present in the ejected gas [as argued e.g. by \citet{dysoetal89}], or whether it emerges in the expanding gas as a consequence of instabilities [as argued e.g. by \citet{capr73} for clumps in PNe]. 

It is difficult to judge whether the clumps reflect inhomogeneities already present early, e.g. in the wind-accelerating zone at a distance of a few stellar radii from the star, or even in the stellar
photosphere or below, as convective cells. We can identify two problems in this context. The clumps ejected from the star in this scenario are likely to have a distribution of properties (mass, size, etc.). If so, one would expect clumps with different properties to reach somewhat different terminal velocities, and once they start to interact with the previous slower wind (which is presumably also clumpy), one would expect clumps of different sizes or masses to accrete different amounts of material from the slow wind, and thus aquire different speeds for the second time. Thus it is hard to explain why the clumps seem to be located 
at equal distances from the star in a thin shell. On the other hand, as a result of the faster wind colliding with a slower wind, instabilities of various types may occur, e.g., Rayleigh-Taylor, Kelvin-Helmholtz, thermal, and gravitational instabilities. It is beyond the scope of this paper to analyse this in detail, but we note that numerical models indicate that structures roughly corresponding to the observed sizes and densities may indeed occur due to Rayleigh-Taylor instabilities \citep{myasetal00}. Once the clump characteristics are determined and their formation mechanism understood, the clump properties can possibly be used as a diagnostic of e.g. stellar evolution.

\begin{acknowledgements}
      This work was supported by the Swedish Research Council and the
      Swedish National Space Agency Board. We are very grateful to an anonymous referee for constructive comments.
\end{acknowledgements}


\end{document}